\def\year{2021}\relax
\documentclass[letterpaper]{article} % DO NOT CHANGE THIS
\usepackage{aaai20}  % DO NOT CHANGE THIS
\usepackage{times}  % DO NOT CHANGE THIS
\usepackage{helvet} % DO NOT CHANGE THIS
\usepackage{courier}  % DO NOT CHANGE THIS
\usepackage[hyphens]{url}  % DO NOT CHANGE THIS
\usepackage{booktabs}       % professional-quality tables
\usepackage{graphicx} % DO NOT CHANGE THIS
\def\UrlFont{\rm}  % DO NOT CHANGE THIS
\usepackage{graphicx}  % DO NOT CHANGE THIS
\usepackage{tabularx,tabulary}
\usepackage{dirtree}
\usepackage{pifont} % Mabye for the Numerical Labels \ding{202} see https://en.wikibooks.org/wiki/LaTeX/Special_Characters
\usepackage{dirtree}
\usepackage{nameref}
\usepackage{setspace}
\usepackage{amsmath}

\usepackage{subfig}

\usepackage[dvipsnames,usenames]{xcolor}

\usepackage{float}

\usepackage{subfig}

\usepackage{enumitem}

\usepackage{booktabs} % For formal tables

\usepackage{csquotes}
\usepackage{amsmath}
\usepackage{outlines}

\usepackage{siunitx}
\usepackage[usenames]{xcolor}

\definecolor{col_anger}{HTML}{D60000}
\definecolor{col_fear}{HTML}{009800}
\definecolor{col_anticipation}{HTML}{FE7F00}
\definecolor{col_surprise}{HTML}{008BE2}
\definecolor{col_joy}{HTML}{FFEA52}
\definecolor{col_sadness}{HTML}{0000CA}
\definecolor{col_trust}{HTML}{52FF52}
\definecolor{col_disgust}{HTML}{E000E0}

\def\capfirstletteraux#1#2\relax{\uppercase{#1}\lowercase{#2}}

\usepackage{marvosym}

\usepackage{graphicx}
\usepackage{graphbox} %loads graphicx package
\usepackage{array}
\newcolumntype{P}[1]{>{\centering\arraybackslash}m{#1}}
\newcolumntype{L}[1]{>{\arraybackslash}m{#1}}
\newcolumntype{R}{>{\raggedleft\arraybackslash}X}

\usepackage{xspace}
\newcommand\ie{i.\,e.\xspace}
\newcommand\eg{e.\,g.\xspace}

\newcommand\US{U.\,S.\xspace}

\newcommand{\mathup}[1]{\mathrm{#1}}

\DeclareMathOperator{\logit}{logit}

\usepackage{url}
\usepackage[normalem]{ulem}
\usepackage{balance} 

\makeatletter
\renewcommand{\fps@figure}{htb}         % default {tbp}
\renewcommand{\fps@table}{htb}         % default {tbp}
\makeatother

\allowdisplaybreaks

\newcommand{\citeeg}[1]{(e.g., \citeauthor{#1} \citeyear{#1})}

\title{Community-Based Fact-Checking on Twitter's Birdwatch Platform}

\author{Nicolas Pröllochs\\
JLU Giessen, Germany\\
nicolas.proellochs@wi.jlug.de}

\begin{document}

\maketitle

\begin{abstract}
Misinformation undermines the credibility of social media and poses significant threats to modern societies. As a countermeasure, Twitter has recently introduced ``Birdwatch,'' a community-driven approach to address misinformation on Twitter. On Birdwatch, users can identify tweets they believe are misleading, write notes that provide context to the tweet and rate the quality of other users' notes. In this work, we empirically analyze how users interact with this new feature. For this purpose, we collect {all} Birdwatch notes and ratings between the introduction of the feature in early 2021 and end of July 2021. We then map each Birdwatch note to the fact-checked tweet using Twitter's historical API. In addition, we use text mining methods to extract content characteristics from the text explanations in the Birdwatch notes (\eg, sentiment). Our empirical analysis yields the following main findings: (i) users more frequently file Birdwatch notes for misleading than not misleading tweets. These misleading tweets are primarily reported because of factual errors, lack of important context, or because they treat unverified claims as facts. (ii) Birdwatch notes are more helpful to other users if they link to trustworthy sources and if they embed a more positive sentiment. (iii) The social influence of the author of the source tweet is associated with differences in the level of user consensus. For influential users with many followers, Birdwatch notes yield a lower level of consensus among users and community-created fact checks are more likely to be seen as being incorrect and argumentative. Altogether, our findings can help social media platforms to formulate guidelines for users on how to write more helpful fact checks. At the same time, our analysis suggests that community-based fact-checking faces challenges regarding opinion speculation and polarization among the user base.
\end{abstract}

%%%
%%% The code below is generated by the tool at http://dl.acm.org/ccs.cfm.
%%% Please copy and paste the code instead of the example below.
%%%
%
%\begin{CCSXML}
%<ccs2012>
%<concept>
%<concept_id>10003120.10003130.10003131.10011761</concept_id>
%<concept_desc>Human-centered computing~Social media</concept_desc>
%<concept_significance>500</concept_significance>
%</concept>
%<concept>
%<concept_id>10003120.10003130.10011762</concept_id>
%<concept_desc>Human-centered computing~Empirical studies in collaborative and social computing</concept_desc>
%<concept_significance>500</concept_significance>
%</concept>
%<concept>
%<concept_id>10010147.10010341.10010342.10010343</concept_id>
%<concept_desc>Computing methodologies~Modeling methodologies</concept_desc>
%<concept_significance>300</concept_significance>
%</concept>
%<concept>
%<concept_id>10010405.10010455.10010461</concept_id>
%<concept_desc>Applied computing~Sociology</concept_desc>
%<concept_significance>100</concept_significance>
%</concept>
%</ccs2012>
%\end{CCSXML}
%
%\ccsdesc[500]{Human-centered computing~Social media}
%\ccsdesc[500]{Human-centered computing~Empirical studies in collaborative and social computing}
%\ccsdesc[300]{Computing methodologies~Modeling methodologies}
%\ccsdesc[100]{Applied computing~Sociology}

%%
%% Keywords. The author(s) should pick words that accurately describe
%% the work being presented. Separate the keywords with commas.
%\keywords{Social media, misinformation, fake news, fact-checking, crowd wisdom, regression analysis}

\maketitle

\section{Introduction}

% Misinformation

Misinformation on social media has become a major focus of public debate and academic research. Previous works have demonstrated that misinformation is widespread on social media platforms and that misinformation diffuses significantly farther, faster, deeper, and more broadly than the truth \citeeg{Vosoughi.2018,Prollochs.2021a,Prollochs.2021b}. Concerns about misinformation on social media have been rising in recent years, particularly given its potential impacts on elections \cite{Aral.2019,Bakshy.2015,Grinberg.2019}, public health \cite{Broniatowski.2018}, and public safety \cite{Starbird.2017}. Major social media providers (\eg, Twitter, Facebook) thus have been called upon to develop effective countermeasures to combat the spread of misinformation on their platforms \cite{Lazer.2018,Pennycook.2021}.

%% RW

A crucial prerequisite to curb the spread of misinformation on social media is its accurate identification \cite{Pennycook.2019}. Predominant approaches to identify misinformation on social media can be grouped into two categories. First, human-based systems in which (human) experts or fact-checking organizations (\eg, \url{snopes.com}, \url{politifact.com}, \url{factcheck.org}) determine the veracity \cite{Hassan.2017,Shao.2016}. Second, machine learning-based systems can automatically classify veracity \cite{Ma.2016,Qazvinian.2011}. Here machine learning models are typically trained to classify misinformation using content-based features (\eg text, images, video), context-based features (\eg, time, location), or based on propagation patterns (\ie, how misinformation circulates among users). Yet both approaches have inherent drawbacks: (i) expert's verification tends to be accurate but is difficult to scale given the limited number of professional fact-checkers \cite{Pennycook.2019}. (ii) Machine learning-based detection is scalable but the prediction performance tends to be unsatisfactory \cite{Wu.2019}. Complementary approaches are thus necessary to identify misinformation on social media both accurately and at scale. %\cite{Moravec.2019}.

% Paragraph crowd-based fact-checked; crowd-sourced fact-checking:

As an alternative, recent research has proposed to build on collective intelligence and the ``wisdom of crowds'' to fact-check social media content \cite{Micallef.2020,Bhuiyan.2020,Pennycook.2019,Epstein.2020,Allen.2020,Allen.2020b,Godel.2021}. The wisdom of crowds is the phenomenon that when many individuals independently make an assessment, the aggregate user judgments will be closer to the truth than most individual estimates or even experts \cite{Frey.2020}. Applying the concept of crowd wisdom to fact-checking is appealing as it would allow for large numbers of fact-checks that can be inexpensively and frequently acquired \cite{Allen.2020b,Pennycook.2019}. 
However, research and earlier attempts to harness the wisdom of crowds to identify misleading social media content have so far produced mixed results.
On the one hand, experimental evidence suggest that the assessment of even small crowds is comparable to those from experts \cite{Allen.2020b,Bhuiyan.2020,Pennycook.2019}. On the other hand, existing crowd-based fact-checking initiatives such as \emph{TruthSquad}, \emph{Factcheck.EU}, and \emph{WikiTribune} had limited success and did not prove to be a model to produce high-quality fact-checks at scale \cite{Bhuiyan.2020}. 
Due to quality issues with crowd-created fact-checks, these initiatives required workarounds involving the assessment of experts -- which again limited their scalability \cite{Bhuiyan.2020,Bakabar.2018}. In sum, it has been found to be challenging to implement real-world community-based fact-checking platforms that uphold high quality and scalability. 

%% Birdwatch

\begin{figure}%[H]
\captionsetup{position=top}
%    \begin{minipage}[b]{0.6\textwidth}
				\centering
				% Requires \usepackage{graphicx}
				%\vspace{-.1cm}
				\subfloat[Source tweet]{\fbox{\includegraphics[width=.85\linewidth]{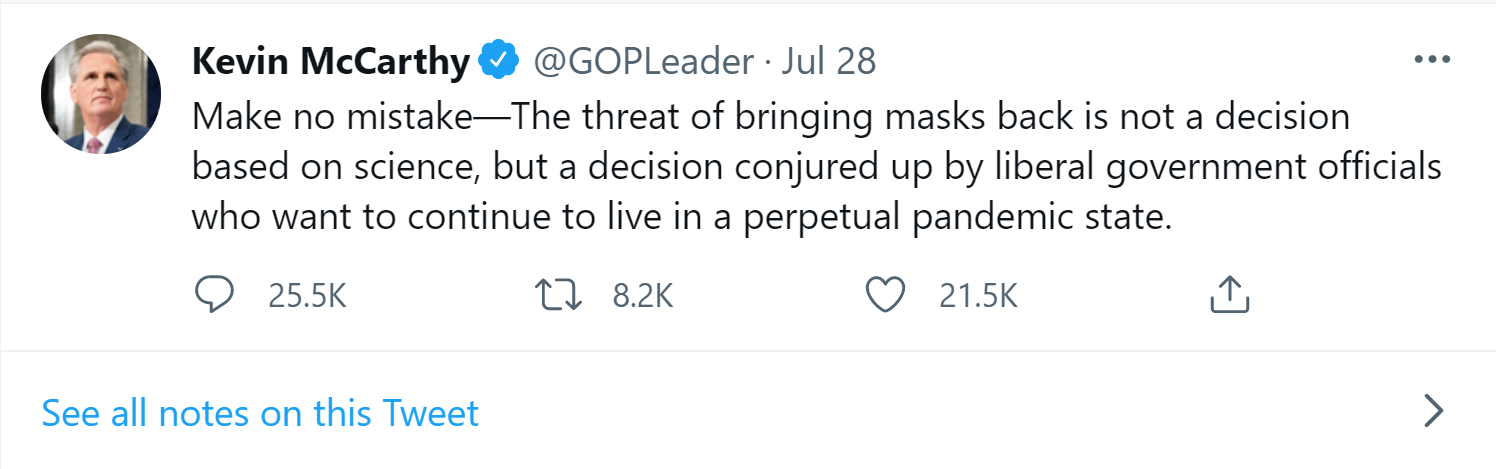}}\label{fig:screenshot_source}}
				\hspace{0.35cm}
				\subfloat[Birdwatch note]{\fbox{\includegraphics[width=.85\linewidth]{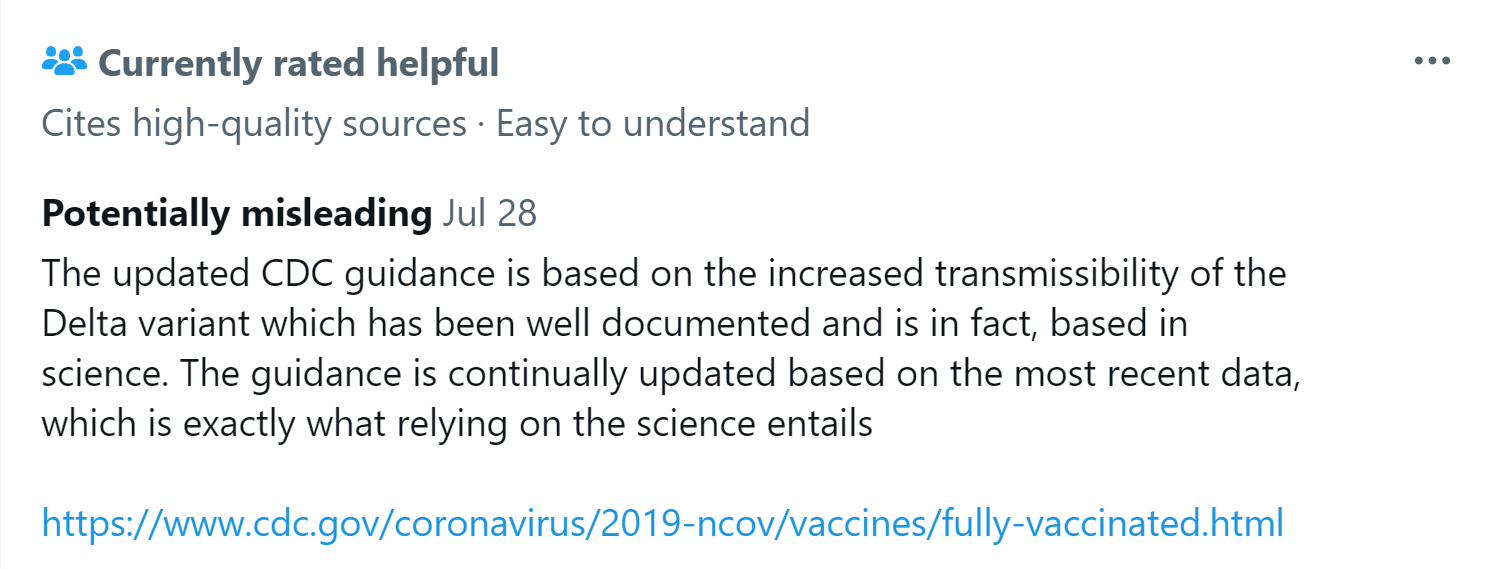}}\label{fig:screenshot_note}}
				\caption{Screenshot of an exemplary community-created fact-check on Birdwatch.}
	\label{fig:screenshot} 
%    \end{minipage}
\end{figure}

Informed by these findings, Twitter has recently launched ``Birdwatch,'' a new attempt to address misinformation on social media by harnessing the wisdom of crowds. Different from earlier crowd-based fact-checking initiatives, Birdwatch is a community-driven approach to identify misleading tweets directly on Twitter (see example in Fig.~\ref{fig:screenshot}). The idea is that the user base on Twitter provides a wealth of knowledge that can help in the fight against misinformation. On Birdwatch, users can identify tweets they believe are misleading (\eg, factual errors) or not misleading (\eg, satire) and write (textual) notes that provide \emph{context} to the tweet. They can also add links to their sources of information. A key feature of Birdwatch is that it implements a rating mechanism that allows users to rate the quality of other participants' notes. These ratings should help to identify the context that people will find most helpful and raise its visibility to other users. While the Birdwatch feature is currently in pilot phase and only directly visible on tweets to pilot participants %via a separate Birdwatch website 
in the \US, Twitter's goal is that Birdwatch will be available to everyone on Twitter. 

%% RQs

\textbf{Research goal: }
In this work, we provide a holistic analysis of the Birdwatch pilot on Twitter. We empirically analyze how users interact with Birdwatch and study factors that make community-created fact checks more likely to be perceived as helpful or unhelpful by other users. Specifically, we address the following research questions: 
\begin{itemize}
\item \textbf{(RQ1)} \emph{What are specific reasons due to which Birdwatch users report tweets?} 
\item \textbf{(RQ2)} \emph{How do Birdwatch notes for tweets categorized as being misleading vs. not misleading differ in terms of their content characteristics (\eg, sentiment, length)?}
\item \textbf{(RQ3)} \emph{Are tweets from Twitter accounts with certain characteristics (\eg, politicians, accounts with many followers) more likely to be fact-checked on Birdwatch?}
\item \textbf{(RQ4)} \emph{Which characteristics of Birdwatch notes are associated with greater helpfulness for other users?}
\item \textbf{(RQ5)} \emph{How is the level of consensus among users associated with the social influence of the author of the source tweet?}
\end{itemize}

\textbf{Methodology:} To address our research questions, we collect \emph{all} Birdwatch notes and ratings from the Birdwatch website between the introduction of the feature on January 23, 2021, and the end of July 2021. This comprehensive dataset contains \num{11802} Birdwatch notes and \num{52981} ratings. We use text mining methods to extract content characteristics from the text explanations of the Birdwatch notes (\eg sentiment, length). In addition, we employ the Twitter historical API to collect information about the source tweets (\ie, the fact-checked tweets) referenced in the Birdwatch notes. We then perform an empirical analysis of the observational data and use regression models to understand how (i) the user categorization, (ii) the content characteristics of the text explanation, and (iii) the social influence of the source tweet are associated with the helpfulness of Birdwatch notes.

\textbf{Contributions:} To the best of our knowledge, this study is the first to present a thorough empirical analysis of Twitter's Birdwatch feature. In contrast to earlier studies focusing on whether crowds are \emph{able} to accurately assess social media content, we contribute to research into misinformation and crowdsourced fact-checking by shedding light on how users \emph{interact} with a community-based fact-checking system. Our work yields the following main findings:
\begin{enumerate}
\item Users file a larger number of Birdwatch notes for misleading than for not misleading tweets. Misleading tweets are primarily reported because of factual errors, lack of important context or because they treat unverified claims as fact. Not misleading tweets are primarily reported because they are perceived as factually correct or because they clearly refer to personal opinion or satire. 
\item Birdwatch notes filed for misleading vs. not misleading tweets differ in terms of their content characteristics. For tweets reported as misleading, authors of Birdwatch notes use a more negative sentiment, more complex language, and tend to write longer text explanations. 
\item Tweets from influential politicians (on both sides of the political spectrum in the \US) are more likely to be fact-checked on Birdwatch. For many of these accounts, Birdwatch users overwhelmingly report misinformation.
\item Birdwatch notes reporting misleading tweets are perceived as being more helpful by other Birdwatch users. 
Birdwatch notes are perceived as being particularly helpful if they provide trustworthy sources and embed a positive sentiment. 
\item The social influence of the author of the source tweet is associated with differences in the level of user consensus. For influential users with many followers, Birdwatch notes receive more total votes (helpful \& unhelpful) but a lower helpfulness ratio (\ie, a lower level of consensus). Also, Birdwatch notes for influential users are particularly likely to be seen as incorrect and argumentative. 
\end{enumerate}

\textbf{Implications:} 
Our findings have direct implications for the Birdwatch platform and future attempts to implement community-based approaches to combat misinformation on social media. We show that users perceive a relatively high share of community-created fact checks as being informative, clear and helpful. Here we find that encouraging users to provide {context} (\eg, by linking trustworthy sources) and avoid the use of inflammatory language is crucial for users to perceive community-based fact checks as helpful. Despite showing promising potential, our analysis suggests that Birdwatch's community-driven approach faces challenges concerning opinion speculation and polarization among the user base -- in particular with regards to influential accounts.

\section{Background}

\subsection{Misinformation on Social Media}

Social media has become a prevalent platform for consuming and sharing information online \cite{Bakshy.2015}. It is estimated that almost 62\% of the adult population consume news via social media, and this proportion is expected to increase further \cite{Pew.2016}. As any user can share information, quality control for the content has essentially moved from trained journalists to regular users \cite{Kim.2019}. The inevitable lack of oversight from experts makes social media vulnerable to the spread of misinformation \cite{Shao.2016}. Social media platforms have indeed been observed to be a medium that disseminates vast amounts of misinformation \cite{Vosoughi.2018}. Several works have studied diffusion characteristics of misinformation on social media \cite{Friggeri.2014,Vosoughi.2018}, finding that misinformation spreads significantly farther, faster, deeper, and more broadly than the truth. The presence of misinformation on social media has detrimental consequences on how opinions are formed and the offline world \cite{Allcott.2017,Bakshy.2015,DelVicario.2016,Oh.2013}. Misinformation thus threatens not only the reputation of individuals and organizations, but also society at large.

% Reasons

Previous research has identified several reasons why misinformation is widespread on social media. On the one hand, misinformation pieces are often intentionally written to mislead other users \cite{Wu.2019}. It is thus difficult for users to spot misinformation from the content itself. On the other hand, the vast majority of social media users do not fact-check articles they read \cite{Geeng.2020,Vo.2018}. This indicates that social media users are often in a hedonic mindset and avoid cognitive reasoning such as verification behavior \cite{Moravec.2019}. A recent study further suggests that the current design of social media platforms may discourage users from reflecting on accuracy \cite{Pennycook.2021}. 

The spread of misinformation can also be seen as a social phenomenon. Online social networks are characterized by homophily \cite{McPherson.2001}, (political) polarization \cite{Levy.2021}, and echo chambers \cite{Barbera.2015}. In these information environments with low content diversity and strong social reinforcement, users tend to selectively consume information that shares similar views or ideologies while disregarding contradictory arguments \cite{Ecker.2010}. These effects can even be exaggerated in the presence of repeated exposure: once misinformation has been absorbed, users are less likely to change their beliefs even when the misinformation is debunked \cite{Pennycook.2018}. 

\subsection{Identification of Misinformation}

%\TODO{Intro paragraph?}

\textbf{Identification via expert's verification:} Misinformation can be manually detected by human experts. In traditional media (\eg, newspapers), this task is typically carried out by professional journalists, editors, and fact-checkers who verify information before it is published \cite{Kim.2019}. %However, this approach does not scale to the volume and speed of content that is generated in social media \cite{Tschiatschek.2018}. %On social media, everyone has the possibility to share information. 
With the possibility for everyone to share information, social media essentially transfers quality control to platform users \cite{Kim.2019}. This has given rise to a number of third-party fact-checking organizations such as \url{snopes.com} and \url{politifact.com} that thoroughly investigate and debunk rumors \cite{Wu.2019,Vosoughi.2018}. The fact-checking assessments are consumed and broadcast by social media users like any other type of news content and are supposed to help users to identify misinformation on social media \cite{Shao.2016}. 
However, the experts' verification approach has inherent drawbacks: (i) it does not scale to the volume and speed of content that is generated in social media. Users can create misleading content at a much faster rate than fact-checkers can evaluate it. Misinformation is thus likely to go unnoticed during its period of peak virality \cite{Epstein.2020}. (ii) Approximately 50\% of Americans overall think that fact-checkers are biased and distrust fact-checking corrections \cite{Poynter.2019}. Hence even when users become aware of the fact-checks of fact-checking organizations, their impact may be limited by lack of trust. In fact, Vosoughi et al. \citeyear{Vosoughi.2018} show that \emph{fact-checked} false rumors are more viral than the truth. 

\textbf{Identification using machine learning methods:} Previous research has intensively focused on the problem of misinformation detection via means of supervised machine learning \cite{Castillo.2011,Ducci.2020,Vosoughi.2017}. Researchers typically collect posts and their labels from social media platforms, and then train a classifier based on diverse sets of features. For instance, {content-based approaches} aim at directly detecting misinformation based on its content, such as text, images, and video. However, different from traditional text categorization tasks, misinformation posts are deliberately made seemingly real and accurate. The predictive power of the actual content in detecting misinformation is thus limited \cite{Wu.2019}. Other common information sources include context-based features (\eg, time, location), or propagation patterns (\ie, how misinformation circulates among users) \citeeg{Kwon.2014}. Yet also with these features, the lack of ground truth labels poses serious challenges when training a machine learning classifier -- in particular in the early stages of the spreading process \cite{Wu.2019}. 

\textbf{Identification through wisdom of crowds:} Recent research has proposed to outsource fact-checking of misinformation to non-expert fact-checkers in the crowd \cite{Micallef.2020,Bhuiyan.2020,Pennycook.2019,Epstein.2020,Allen.2020,Allen.2020b}. The idea is to identify misleading social media content by harnessing the wisdom of crowds \cite{Woolley.2010}. The wisdom of crowds has been repeatedly observed in a wide range of settings, including online platforms such as Wikipedia and Stack Overflow, where the crowd ensures relatively trustworthy and high-quality accumulation of knowledge \cite{Okoli.2014}. Applying the concept of crowd wisdom to fact-checking on social media may have crucial advantages: (i) compared to the expert's verification approach, which is limited by the number of professional fact-checkers, crowd-based approaches allow to identify misinformation at a large scale \cite{Pennycook.2019}. (ii) Community-based fact-checking addresses the problem that many users distrust professional fact-checks \cite{Poynter.2019}. (iii) The wisdom of crowds literature suggests that, even if the ratings of individual users are noisy and ineffective, in the aggregate user judgments may be highly accurate \cite{Woolley.2010}. Experimental studies found that the crowd can in fact be quite accurate in identifying misleading social media content. Here the assessment of even relatively small crowds is comparable to those from experts \cite{Bhuiyan.2020,Epstein.2020,Pennycook.2019}. 

%% Challenges
Yet, while the crowd may be generally \emph{able} to accurately identify misinformation, not all users may always \emph{choose} to do so \cite{Epstein.2020}. Important challenges include %lack of knowledge, 
manipulation attempts \cite{Luca.2016}, lack of engagement in cognitive reasoning \cite{Pennycook.2019b}, and politically motivated reasoning \cite{Kahan.2017}. Each of these factors may hinder an effective fact-checking system. For example, users could purposely try to game the fact-checking system by reporting social media posts as being misleading that do not align with their ideology (irrespective of the perceived veracity) or to achieve partisan ends \cite{Luca.2016}. Furthermore, the high level of (political) polarization of social media users \cite{Conover.2011,Barbera.2015}, can result in vastly different interpretations of facts or even entirely different sets of acknowledged facts \cite{Otala.2021}. Earlier crowd-based fact-checking initiatives such as \emph{TruthSquad}, \emph{Factcheck.EU}, and \emph{WikiTribune} \cite{Oriordan.2019,Florin.2010} indeed experienced significant quality-issues with community-based fact-checks \cite{Bhuiyan.2020,Bakabar.2018}. These initiatives required workarounds or hybrid approaches involving final judgments by experts or delegating primary research to experts and secondary tasks to the crowd \cite{Bhuiyan.2020,Bakabar.2018}. This suggests that it is challenging to implement real-world community-based fact-checking platforms that uphold high quality and scalability. 
More precisely, as pointed out by \citeauthor{Epstein.2020}~(\citeyear{Epstein.2020}), the challenges that must be overcome in order to implement it successfully are social in nature rather than technical, and thus involve empirical questions about how people \emph{interact} with community-based fact-checking systems. 

\begin{table*}
\renewcommand{\baselinestretch}{.8} 
\scriptsize
\def\UrlFont{\em}
\begin{tabular}{lp{0.38\textwidth}p{0.09\textwidth}p{0.275\textwidth}ll}
\toprule
\multicolumn{2}{l}{\textbf{Source tweet}} & \multicolumn{2}{l}{\textbf{Birdwatch note}} & \multicolumn{2}{l}{\textbf{Ratings}}\\
\cmidrule(lr){3-4} \cmidrule(lr){5-6}
 && \textbf{Categorization} &  \textbf{Text explanation} & \textbf{Helpful} & \textbf{Unhelpful} \\% & $\boldsymbol{\mathit{HVotes}}$ & $\boldsymbol{\mathit{UVotes}}$ \\
\midrule
\#1& \textbf{Paul Elliott Johnson (\emph{@RhetoricPJ})}: ``\texttt{Rush Limbaugh had a regular radio segment where he would read off the names of gay people who died of AIDS and celebrate it and play horns and bells and stuff.}'' & \uwave{Misleading} & ``\texttt{These segments are fictitious and have never happened.}'' & 3 & 21 \\
\addlinespace
\#1& \textbf{Alexandria Ocasio-Cortez (\emph{@AOC})}: ``\texttt{I am happy to work with Republicans on this issue where there’s common ground, but you almost had me murdered 3 weeks ago so you can sit this one out. Happy to work w/ almost any other GOP that aren’t trying to get me killed. In the meantime if you want to help, you can resign. [LINK]}''  & {Misleading} & ``\texttt{While I can understand AOC being worried she may have been hurt by the Capitol Protesters she cannot claim Ted Cruz tried to have her killed without providing proof. That's dangerous misinformation by falsely accusing Ted Cruz of a serious crime.}'' & 47 & 12 \\
\addlinespace
\#3& \textbf{Adam Kinzinger (\emph{@RepKinzinger})}: ``\texttt{The vast majority of Americans believe in universal background checks. As a gun owner myself, I firmly support the Second Amendment but I also believe we have to be willing to make some changes for the greater good.  Read my full statement on \#HR8 here: [LINK]}'' & Not misleading & ``\texttt{Universal background checks enjoy high levels of public support; a 2016 representative survey found 86\% of registered voters in the United States supported the measure. [LINK]}'' & 7 & 0 \\%& \colorbox{RoyalBlue!15}{Not misleading} \url{https://www.snopes.com/fact-check/rush-limbaugh-mock-aids-gays/}\\
\addlinespace
\#4& \textbf{Richard Grenell (\emph{@RichardGrenell})}: ``\texttt{Congratulations to @PeteButtigieg on becoming the second openly gay member of a President’s Cabinet.  Welcome to the club! [LINK]}'' & Not misleading & ``\texttt{This is correct. Ric Grenell was the first openly gay cabinet member. [LINK]}'' & 5 & 2 \\
\bottomrule 
\end{tabular}
\caption{Examples of Birdwatch notes and ratings. The columns ``Helpful'' and ``Unhelpful'' refer to the number of users who have responded ``Yes'' (``No'') to the rating question ``Is this note helpful?'' 
Wavy underlining indicates fact-checks that are clearly false (\ie, abuse of the fact-checking feature).}
\label{tbl:examples}
\end{table*}

\section{What is Birdwatch?}

On January 23, 2021, Twitter has launched the ``Birdwatch'' feature, a new approach to address misinformation on social media by harnessing the wisdom of crowds. In contrast to earlier crowd-based initiatives to fact-checking, Birdwatch is a community-driven approach to identify misleading tweets directly on Twitter. On Birdwatch, users can identify tweets they believe are misleading and write (textual) notes that provide \emph{context} to the tweet. Birdwatch also features a rating mechanism that allows users to rate the quality of other users' notes. These ratings are supposed to help to identify the context that people will find most helpful and raise its visibility to other users. The Birdwatch feature is currently in the pilot phase in the \US and only pilot participants can see community-written fact-checks directly on tweets when browsing Twitter. Users not participating in the pilot can access Birdwatch via a separate Birdwatch website\footnote{The Birdwatch website is available via \url{birdwatch.twitter.com}}. Twitter's goal is that community-written notes will be visible directly on tweets, available to everyone on Twitter.

\textbf{Birdwatch notes:} 
Users can add Birdwatch notes to any tweet they come across and think might be misleading. Notes are composed of (i) multiple-choice questions that allow users to state why a tweet might or might not be misleading; (ii) an open text field where users can explain their judgment, as well as link to relevant sources. The maximum number of characters in the text field is 280. Here, each URL counts as only 1 character towards the 280 character limit. After it's submitted, the note is available for other users to read and rate. Birdwatch notes are public, and anyone can browse the Birdwatch website to see them. 

\textbf{Ratings:} 
Users on Birdwatch can rate the helpfulness of notes from other users. These ratings are supposed to help to identify which notes are most helpful and to allow Birdwatch to raise the visibility of the context that is found most helpful by a wide range of contributors. Tweets with Birdwatch notes are highlighted with a Birdwatch icon that helps users to find them. Users can then click on the icon to read all notes others have written about that tweet and rate their quality (\ie, whether or not the Birdwatch note is helpful). Notes rated by the community to be particularly helpful then receive a currently rated helpful badge. 

\textbf{Illustrative examples:} Tbl.~\ref{tbl:examples} presents four examples of Birdwatch notes and helpfulness ratings. The table shows that many Birdwatch notes report tweets that refer to a political topic (\eg, voting rights). We also see that the text explanations differ regarding their informativeness. Some text explanations are longer and link to additional sources that are supposed to support the comments of the author. There are also users trying to abuse Birdwatch by falsely asserting that a tweet is misleading. In such situations, other users can down-vote the note through Birdwatch's rating system. As an example, the (false) assertion in Birdwatch note \#1 received 3 helpful and 21 unhelpful votes. In some cases, users also fact-check tweets for which veracity may be seen as being difficult to assess. For instance, Birdwatch note \#2 addresses the source tweet from Democratic Congresswoman Alexandria Alexandria Ocasio-Cortez as a factual assertion that members of the GOP were trying to get her killed during the Capitol riots on January 6, 2021. However, one may also argue that the tweet can be seen a hyperbolic claim to emphasize that she believes that certain policy views or rhetorical stance encouraged the Capitol rioters. This is reflected in mixed helpfulness ratings: Birdwatch note \#2 received 47 helpful and 12 unhelpful votes.

\section{Data}

\subsection{Data Collection}

We downloaded \emph{all} Birdwatch notes and ratings between the introduction of the feature on January 23, 2021, and the end of July 2021 from the Birdwatch website. This comprehensive dataset contains a total number of \num{11802} Birdwatch notes and \num{52981} ratings. On Birdwatch, multiple users can write Birdwatch notes for the same tweet. The average number of Birdwatch notes per fact-checked tweet is \num{1.31}. Each Birdwatch note has a unique id (\emph{noteId}) and each rating refers to a single Birdwatch note. We merged the ratings with the Birdwatch notes using this \emph{noteId} field. The result is a single dataframe in which each Birdwatch note corresponds to one observation for which we know the number of helpful and unhelpful votes from the rating data.

\subsection{Key Variables}

We now present the key variables that we extracted from the Birdwatch data. All of these variables will be empirically analyzed in the next sections: 

\begin{itemize}
\item \emph{Misleading:} A binary indicator of whether a tweet has been reported as being misleading by the author of the Birdwatch note ($=1$; otherwise $=0$). 
\item \emph{Text explanation:} The user-entered text explanation (max 280 characters) to explain why a tweet is misleading or not misleading. 
\item \emph{Trustworthy sources:} A binary indicator of whether the author of the Birdwatch note has responded ``Yes'' to the question ``Did you link to sources you believe most people would consider trustworthy?'' ($=1$; otherwise $=0$).
\item \emph{HVotes:} The number of other users who have responded ``Yes'' to the rating question  ``Is this note helpful?'' 
\item \emph{Votes:} The total number of ratings a Birdwatch note has received from other users (helpful \& unhelpful).
\end{itemize}

\subsection{Content Characteristics}

We use text mining methods to extract the following content characteristics from the text explanations of the Birdwatch notes:
\begin{itemize}
\item \emph{Sentiment:} We calculate a sentiment score that measures the extent of positive vs. negative emotions in Birdwatch notes. Our computation follows a dictionary-based approach as in \citeauthor{Vosoughi.2018}~(\citeyear{Vosoughi.2018}). Here we use the NRC emotion lexicon \cite{Mohammad.2013}, which classifies English words into positive and negative emotions. The fraction of words in the text explanations related to positive and negative emotions is then aggregated and averaged to create a vector of positive and negative emotion weights that sum to one. The sentiment score is then defined as the difference between positive and negative emotion scores. 
\item \emph{Text complexity:} We calculate a text complexity score, specifically, the Gunning-Fog index \cite{Gunning.1968}. This index estimates the years of formal education necessary for a person to understand a text upon reading it for the first time: $0.4 \times (\mathit{ASL} + 100 \times n_{\text{wsy} \geq 3} / n_{\text{w}})$, where $\mathit{ASL}$ is the average sentence length (number of words), $n_{\text{w}}$ is the total number of words, and $n_{\text{wsy} \geq 3}$ is the number of words with three syllables or more. A higher value thus indicates greater complexity. 
\item \emph{Word count:} We determine the length of the text explanations in Birdwatch notes as given by the number of words.
\end{itemize}

Our text mining pipeline is implemented in R~4.0.2 using the packages \texttt{quanteda} \cite{Benoit.2018} in Version 2.0.1 and \texttt{sentimentr} \cite{Rinker.2019} in Version 2.7.1. 

\subsection{Twitter Historical API}

Each Birdwatch note addresses a single tweet, \ie, the tweet that has been categorized as being misleading or not misleading by the author of the Birdwatch note. We used the Twitter historical API to map the \emph{tweetID} referenced in each Birdwatch note to the source tweet and collected the following information for the author of each source tweet: 
\begin{itemize}
\item \emph{Account name:} The name of the Twitter account for which the Birdwatch note has been reported.
\item \emph{Followers:} The number of followers, \ie, the number of accounts that follow the author of the source tweet. %on Twitter.
\item \emph{Followees:} The number of followees, \ie, the number of accounts whom the author of the source tweet follows. %on Twitter.
\item \emph{Account age:} The age of the author of the source tweet's account (in years).
\item \emph{Verified:} A binary dummy indicating whether the account of the source tweet has been officially verified by Twitter ($=1$; otherwise $=0$).
\end{itemize}

\section{Empirical Analysis}

\subsection{Analysis of Birdwatch Notes (RQ1 \& RQ2)}

In this section, we analyze how users categorize misleading and not misleading tweets in Birdwatch notes. We also explore the reasons because of which Birdwatch users report tweets and how Birdwatch notes for misleading vs. not misleading differ in terms of their content characteristics.

\begin{figure}%[H]
%\captionsetup{position=top}
%    \begin{minipage}[b]{0.6\textwidth}
				\centering
				% Requires \usepackage{graphicx}
				%\vspace{-.1cm}
				{\includegraphics[width=.7\linewidth]{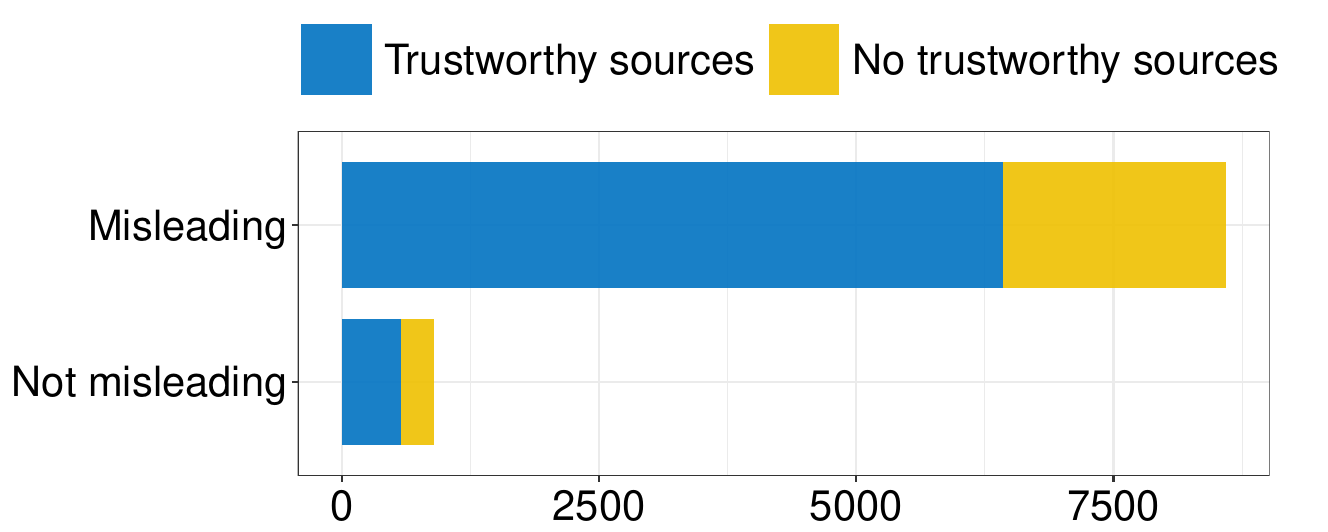}}
				%\hspace{0.35cm}
				%\subfloat[Not helpful]{\includegraphics[width=.22\linewidth]{figures/barplot_helpfulness_notMisleading}\label{fig:structural_virality_ccdf}}
				%\vspace{-.1cm}
				\caption{Number of users who responded ``Yes'' to the question ``Did you link to sources you believe most people would consider trustworthy?''}
				\label{fig:trustworthy} 
%    \end{minipage}
\end{figure}

\textbf{Categorization of tweets in Birdwatch notes:} Fig.~\ref{fig:trustworthy} shows that users file a larger number of Birdwatch notes for misleading than for not misleading tweets. Out of all Birdwatch notes, \SI{90}{\percent} refer to misleading tweets, whereas the remaining \SI{10}{\percent} refer to not misleading tweets. Fig.~\ref{fig:trustworthy} further suggests that authors of Birdwatch notes are more likely to report that they have linked to trustworthy sources for misleading (\SI{75}{\percent}) than for not misleading tweets (\SI{64}{\percent}). 

\begin{figure}%[H]
	\centering
	{\includegraphics[width=\linewidth]{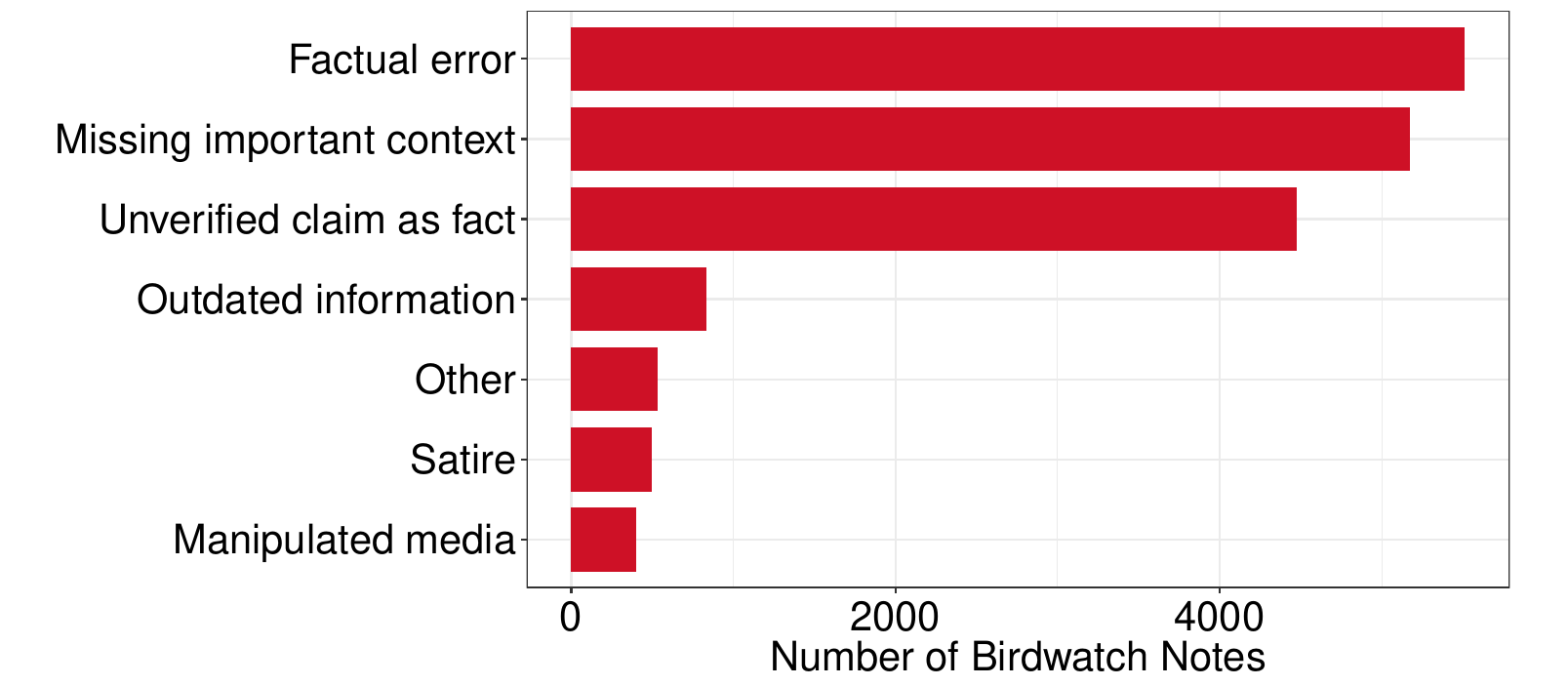}}
	\caption{Number of Birdwatch notes per checkbox answer option in response to the question ``Why do you believe this tweet may be misleading?''}
	\label{fig:ccdf_notes_misleading} 
\end{figure}

\textbf{Why do users report misleading tweets?} Authors of Birdwatch notes also need to answer a checkbox question (multiple selections possible) on why they perceive a tweet as being misleading. Fig.~\ref{fig:ccdf_notes_misleading} shows that misleading tweets are primarily reported because of factual errors (\SI{32}{\percent}), lack of important context (\SI{30}{\percent}), or because they treat unverified claims as facts (\SI{26}{\percent}). Birdwatch notes reporting outdated information (\SI{5}{\percent}), satire (\SI{3}{\percent}), or manipulated media (\SI{2}{\percent}) are relatively rare. Only \SI{3}{\percent} of Birdwatch notes are reported because of other reasons. 

\begin{figure}%[H]
%\captionsetup{position=top}
%    \begin{minipage}[b]{0.6\textwidth}
				\centering
				% Requires \usepackage{graphicx}
				%\vspace{-.1cm}
				{\includegraphics[width=\linewidth]{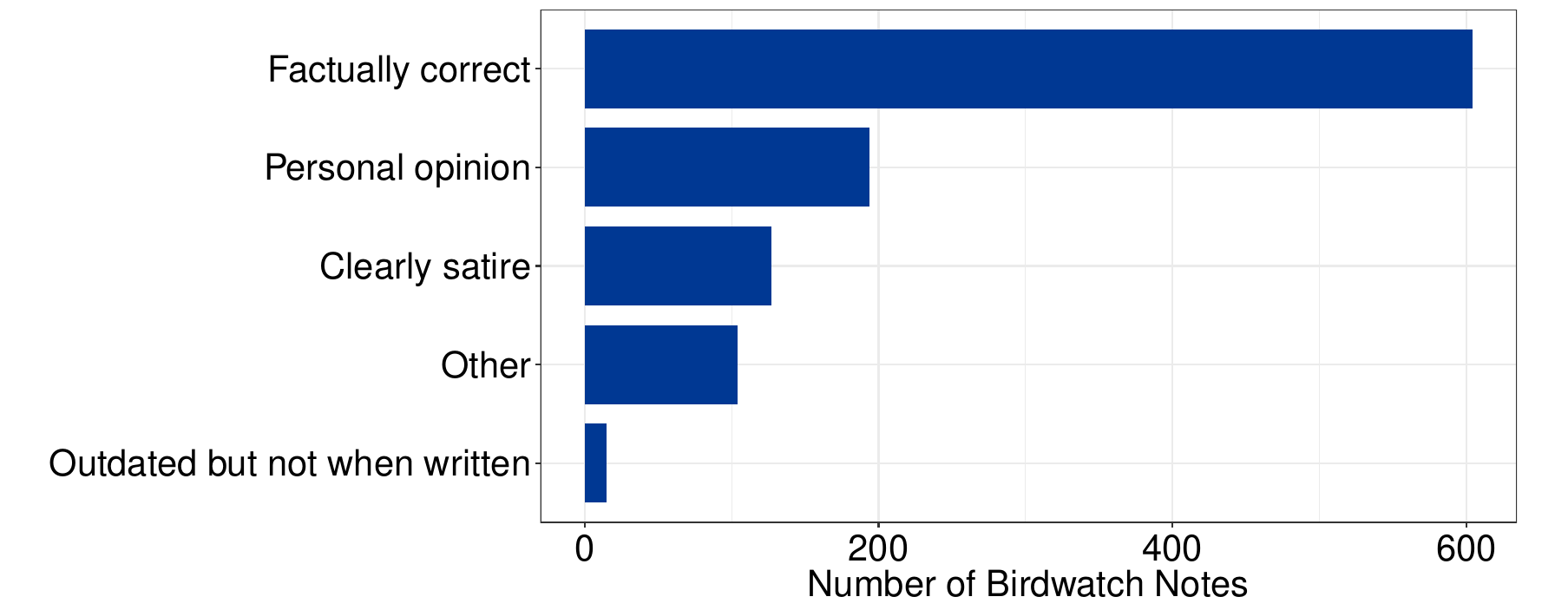}\label{fig:cascade_size_ccdf}}
				%\vspace{-.1cm}
				\caption{Number of Birdwatch notes per checkbox answer option in response to the question ``Why do you believe this tweet is not misleading?''}
	\label{fig:ccdf_notes_notMisleading} 
%    \end{minipage}
\end{figure}

\textbf{Why do users report not misleading tweets?} Fig.~\ref{fig:ccdf_notes_notMisleading} shows that not misleading tweets are primarily reported because they are perceived as factually correct (\SI{58}{\percent}), clearly refer to personal opinion (\SI{19}{\percent}), or satire (\SI{12}{\percent}). Birdwatch users only rarely report tweets containing information that was correct at the time of writing but is now outdated (\SI{1}{\percent}). \SI{10}{\percent} of tweets are reported because of other reasons.

\begin{figure}%[H]
\captionsetup{position=top}
%    \begin{minipage}[b]{0.6\textwidth}
				\centering
				% Requires \usepackage{graphicx}
				%\vspace{-.1cm}
				\subfloat[Sentiment]{\includegraphics[width=.475\linewidth]{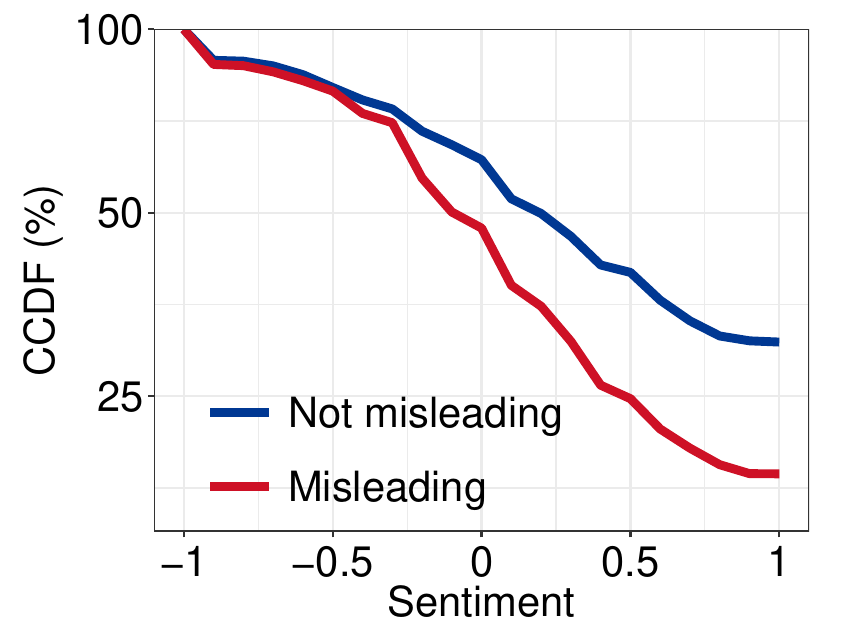}}
				\hspace{0.1cm}
				\subfloat[Text complexity]{\includegraphics[width=.475\linewidth]{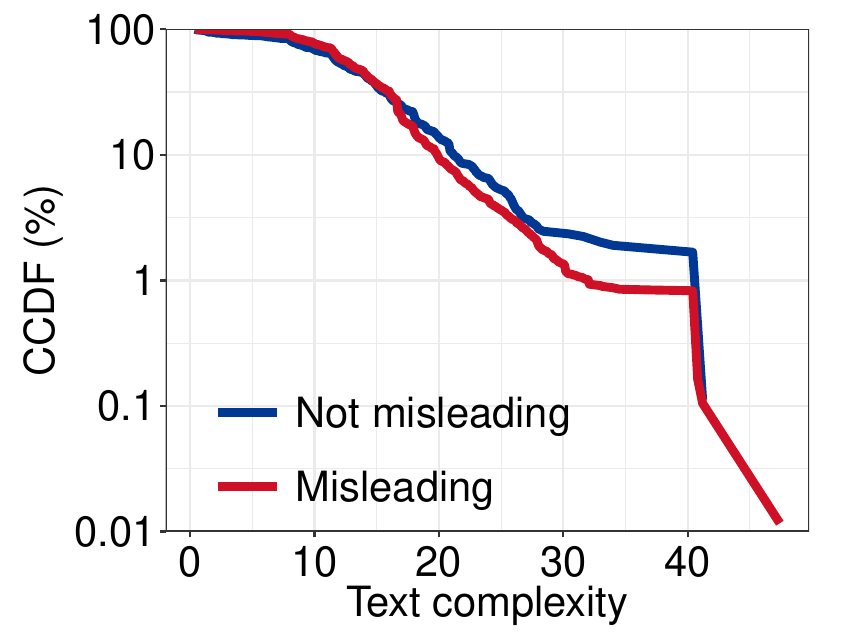}}\\
				%\hspace{0.35cm}
				\subfloat[Word count]{\includegraphics[width=.475\linewidth]{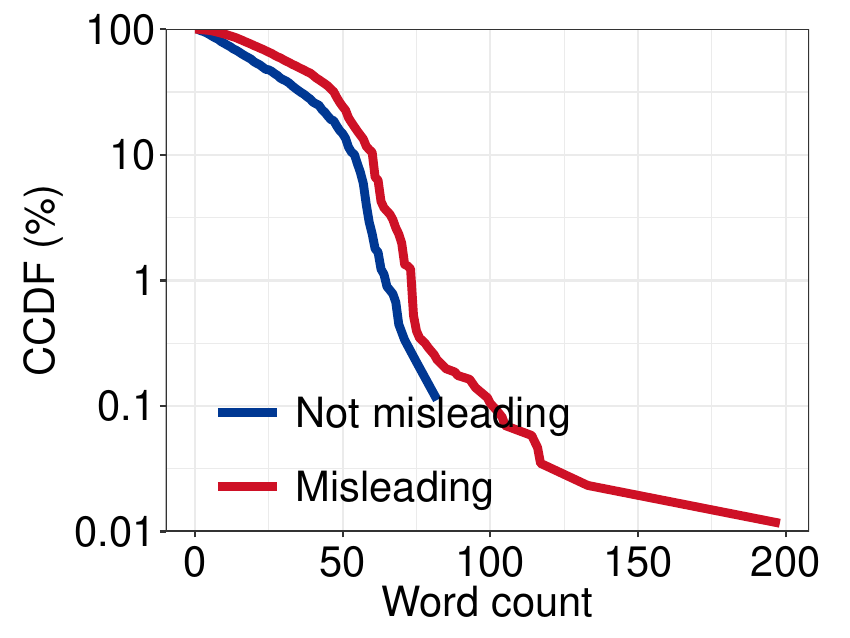}}
				%\hspace{0.35cm}
				%\subfloat[Not trustworthy vs. trustworthy]{\includegraphics[width=.3\linewidth]{figures/ratioHelpful_classification_trustworthySources}\label{fig:structural_virality_ccdf}}
				%\vspace{-.1cm}
				\caption{CCDFs for (a) sentiment, (b) text complexity, and (c) word count in text explanations of Birdwatch notes.}
	\label{fig:ccdf_notes_content} 
%    \end{minipage}
\end{figure}

\textbf{Content characteristics:} The complementary cumulative distribution functions (CCDFs) in Fig.~\ref{fig:ccdf_notes_content} visualize how Birdwatch notes for misleading vs. not misleading tweets differ in terms of their content characteristics. We find that Birdwatch notes reporting misleading tweets use similarly complex language but embed a higher proportion of negative emotions than tweets reported as being not misleading. For tweets reported as being misleading, authors of Birdwatch notes also tend to write longer text explanations.

\subsection{Analysis of Source Tweets (RQ3)}

We now explore whether tweets from Twitter accounts with certain characteristics (\eg, politicians, accounts with many followers) are more likely to be fact-checked on Birdwatch. 

\textbf{Most reported accounts:} Tbl.~\ref{tbl:accounts_notes} reports the names of the ten\footnote{An extended list is provided in the supplements.} Twitter accounts that have gathered the most Birdwatch notes. %In addition, Tbl.~\ref{tbl:accounts_proportion_misleading} reports the names of the ten Twitter accounts with the highest share of fact-checked tweets out of all tweets since the introduction of Birdwatch. Here we focus on accounts that have posted at least 20 tweets during our observation period. 
The table also reports the number of unique users that have filed Birdwatch notes per account. 
We find that many of the most-reported accounts belong to influential politicians from both sides of the political spectrum in the U.~S. -- for which Birdwatch users overwhelmingly report misinformation. For instance, \num{160} Birdwatch notes have been reported for the account of Republican \US House Representantive Marjorie Taylor Greene (\emph{@mtgreenee}), out of which \SI{97}{\percent} have been categorized as being misleading tweets. Similarly, \num{160} Birdwatch notes have been reported for Democratic Congresswoman Alexandria Ocasio-Cortez (\emph{AOC}), out of which \SI{90}{\percent} have been categorized as being misleading tweets. % The only Twitter account in the list in Tbl.~\ref{tbl:accounts_notes} for which the majority of tweets have been reported as being not misleading is the account of Steven Crowder (\emph{scrowder}), an American-Canadian conservative political commentator, media host, and comedian.
The highest number of Birdwatch notes has been filed for \emph{@Quakeprediction}, an account that regularly publishes earthquake predictions. Notably, this account has only been fact-checked by three different users that have filed an extensive number of fact-checks. We further observe many fact-checks for newspapers (\emph{@thehill}) and other prominent public figures such as NYT best-selling author Candace Owens (\emph{@RealCandanceO}).
%The table also shows that some accounts have been reported by only few users whereas for tweets from other accounts are fact-checked my many users.

\begin{table}%[ht]
\scriptsize
\setlength{\tabcolsep}{3pt}
\centering
\begin{tabularx}{\columnwidth}{%>{\em}
p{.52\columnwidth}RRR}
  \toprule
\textbf{{Account name}} &  \textbf{\#Notes}  & \textbf{Misleading} &  \textbf{\#Users} \\ 
  \midrule
	EarthquakePrediction \emph{(@Quakeprediction)} & 257 & 100.00\% &   3 \\ 
  Marjorie Taylor Greene \emph{(@mtgreenee)} & 180 & 96.67\% &  93 \\ 
  Alexandria Ocasio-Cortez\emph{(@AOC)} & 160 & 90.00\% & 105 \\ 
  Lauren Boebert \emph{(@laurenboebert)} & 108 & 91.67\% &  62 \\ 
  Alex Berenson \emph{(@AlexBerenson)} &  78 & 97.44\% &  44 \\ 
  President Biden \emph{(@POTUS)}$^\dagger$ &  74 & 81.08\% &  58 \\ 
  The Hill \emph{(@thehill)} &  66 & 95.45\% &  50 \\ 
  %Dion Cini \emph{(@dioncini)} &  61 & 100.00 &   3 \\ 
  Candace Owens \emph{(@RealCandaceO)} &  49 & 89.80\% &  46 \\ 
  Japan Earthquakes \emph{(@earthquakejapan)} &  44 & 100.00\% &   1 \\ 
  Aaron Rupar \emph{(@atrupar)} &  39 & 84.62\% &  27 \\
	%Rep. Marjorie Taylor Greene (\emph{@mtgreenee}) & 180 & 96.67~\% \\ 
  %Alexandria Ocasio-Cortez (\emph{@AOC}) & 160 & 90.00~\% \\ 
  %Lauren Boebert (\emph{@laurenboebert}) & 108 & 91.67~\% \\ 
  %Alex Berenson (\emph{@AlexBerenson}) &  78 & 97.44~\% \\ 
  %President Biden (\emph{@POTUS}) &  74 & 81.08~\% \\ 
  %The Hill (\emph{@thehill}) &  66 & 95.45~\% \\ 
  %Candace Owens (\emph{@RealCandaceO}) &  49 & 89.80~\% \\ 
  %Aaron Rupar (\emph{@atrupar}) &  39 & 84.62~\% \\ 
  %Dubious Poso (\emph{@JackPosobiec}) &  38 & 84.21~\% \\ 
  %Senator Ted Cruz (\emph{@SenTedCruz}) &  36 & 83.33~\% \\ 
\bottomrule
\multicolumn{4}{p{\columnwidth}}{\footnotesize $^\dagger$The presidential Twitter account \emph{@POTUS} was held by Joe Biden throughout the entire study period. The account has been reset after the transition to the new administration on January 20, 2021.}\\
\end{tabularx}
\caption{Top-10 Twitter accounts with the highest number of Birdwatch notes.% since the introduction of Birdwatch.
}
\label{tbl:accounts_notes}
\end{table}

Next, we explore Twitter accounts with the highest \emph{share} of tweets reported as being misleading (Tbl.~\ref{tbl:accounts_proportion_misleading}). For this purpose, we employed the Twitter historical API to obtain the number of tweets posted by each fact-checked account (\ie, the user timelines) since the introduction of Birdwatch. To identify accounts that received interest by many fact-checkers, we focus on accounts that have been fact-checked by at least 20 unique users. We again find that Birdwatch users have pronounced interest in debunking posts from politicians. The highest share of misleading tweets is found for the account of Alexandria Ocasio-Cortez, for which approximately \SI{6}{\percent} of all posted tweets have been reported as being misleading. Note that Marjorie Taylor Greene appears in the top-10 list with both her private (\emph{@mtgreenee}) and her official governmental account (\emph{@RepMTG}). %In sum, our results suggests that Birdwatch users have pronounced interest in debunking posts from politicians. 

\begin{table}%[ht]
\scriptsize
\setlength{\tabcolsep}{3pt}
\centering
\begin{tabularx}{\columnwidth}{%>{\em}
p{.5\columnwidth}RRR}
  \toprule
\textbf{{Account name}} &  \textbf{\#Posts} & \textbf{Misleading}  &  \textbf{\#Users}\\ 
  \midrule
	Alexandria Ocasio-Cortez \emph{(@AOC)} & 640 & 5.62\% & 105 \\ 
  Rep. Marjorie Taylor Greene \emph{(@RepMTG)} & 303 & 5.61\% &  22 \\ 
  Marjorie Taylor Greene \emph{(@mtgreenee)} & 2182 & 4.90\% &  93 \\ 
  Lauren Boebert \emph{(@laurenboebert)} & 1486 & 4.31\% &  62 \\ 
  Candace Owens \emph{(@RealCandaceO)} & 576 & 3.82\% &  46 \\ 
  President Biden \emph{(@POTUS)} & 1176 & 3.32\% &  58 \\ 
  Alex Berenson \emph{(@AlexBerenson)}$^\dagger$ & 2699 & 1.96\% &  44 \\ 
  Rep. Jim Jordan \emph{(@Jim Jordan)} & 1273 & 1.81\% &  25 \\ 
  Cori Bush \emph{(@CoriBush)} & 414 & 1.69\% &  27 \\ 
  Charlie Kirk \emph{(@charliekirk11)} & 1051 & 1.43\% &  21 \\ 
%Alexandria Ocasio-Cortez (\emph{@AOC}) & 640 & 5.62~\% \\ 
  %Rep. Marjorie Taylor Greene (\emph{@RepMTG}) & 303 & 5.61~\% \\ 
  %Lauren Boebert (\emph{@laurenboebert}) & 1486 & 4.31~\% \\ 
  %Candace Owens (\emph{@RealCandaceO}) & 576 & 3.82~\% \\ 
  %Alex Berenson (\emph{@AlexBerenson}) & 2699 & 1.96~\% \\ 
  %Rep. Jim Jordan (\emph{@JimJordan})$^\dagger$ & 1273 & 1.81~\% \\ 
  %Cori Bush (\emph{@CoriBush}) & 414 & 1.69~\% \\ 
  %Charlie Kirk (\emph{@charliekirk11}) & 1051 & 1.43~\% \\ 
  %Senator Ted Cruz (\emph{@SenTedCruz}) & 1715 & 1.17~\% \\ 
  %Rudy W. Giuliani (\emph{@RudyGiuliani}) & 979 & 0.92~\% \\ 
\bottomrule
%\multicolumn{4}{p{\columnwidth}}{\footnotesize $^\dagger$The presidential twitter account \emph{@POTUS} was held by Joe Biden throughout the entire study period. The account has been resetted after the transition to the new administration on January 20, 2021.}\\
\multicolumn{4}{p{\columnwidth}}{\footnotesize $^\dagger$Account has been suspended by Twitter.}\\
\end{tabularx}
\caption{{Top-10 Twitter accounts with the highest share of misleading tweets. Here we focus on accounts that have been fact-checked by at least 20 different users.}
}
\label{tbl:accounts_proportion_misleading}
\end{table}

\begin{figure}%[H]
\captionsetup{position=top}
%    \begin{minipage}[b]{0.6\textwidth}
				\centering
				% Requires \usepackage{graphicx}
				%\vspace{-.1cm}
				\subfloat[Followers]{\includegraphics[width=.475\linewidth]{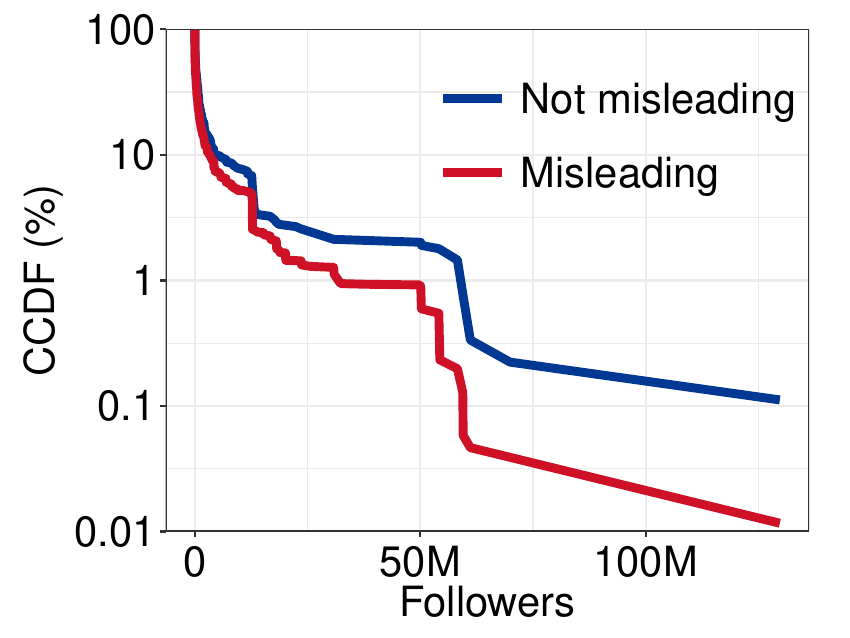}}
				\hspace{0.1cm}
				\subfloat[Followees]{\includegraphics[width=.475\linewidth]{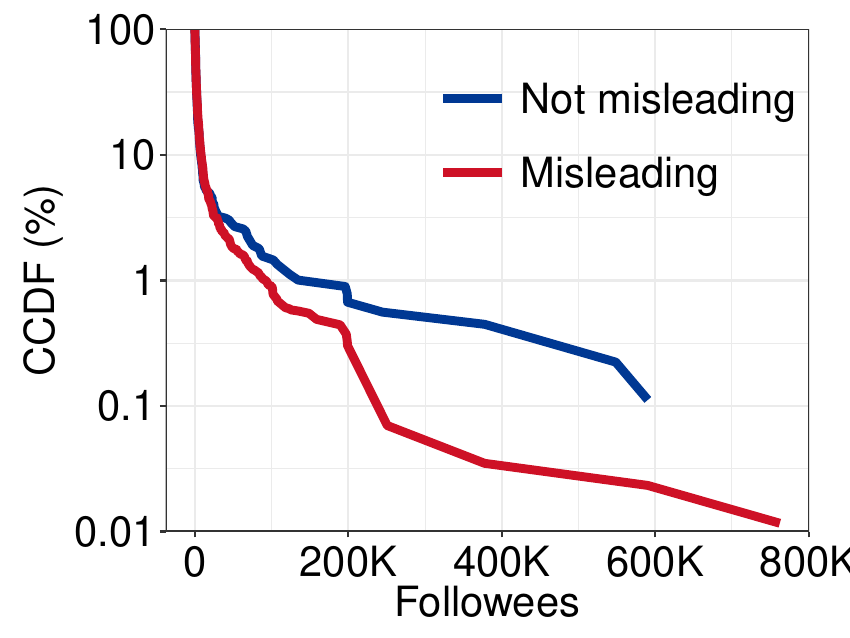}}\\
				%\hspace{0.35cm}
				%\hspace{0.2cm}
				\subfloat[Account age]{\includegraphics[width=.475\linewidth]{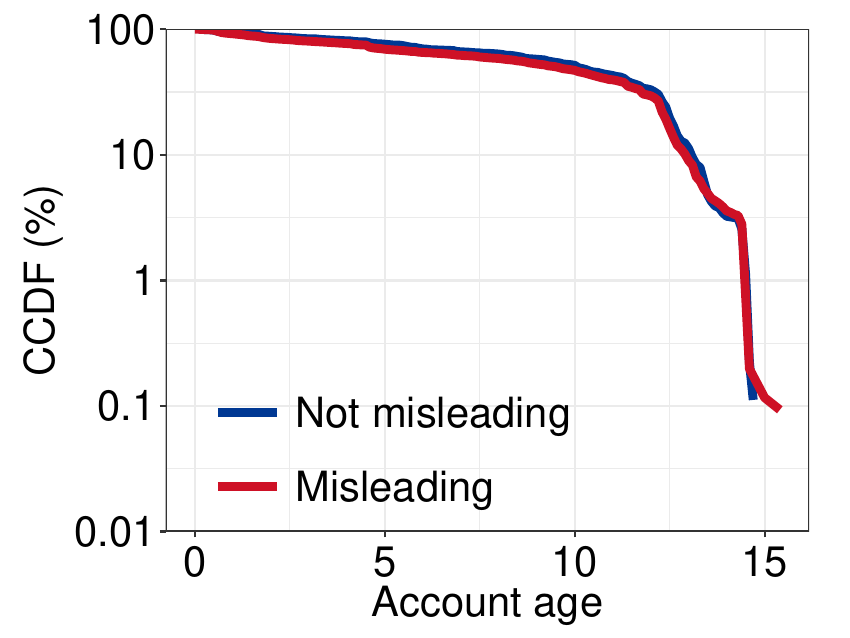}}
				%\hspace{0.35cm}
				%\hspace{0.2cm}
				%\subfloat[Fact-checking delay]{\includegraphics[width=.22\linewidth]{figures/ccdf_source_tweet_diff_factcheck_days}}
				%\subfloat[Verified]{\includegraphics[width=.45\linewidth]{figures/barplot_verified}\label{fig:structural_virality_ccdf}}
				%\vspace{-.1cm}
				\caption{CCDFs for (a) followers, (b) followees, and (c) account age (in years).}%, and (d) difference between timestamps of fact-checks (Birdwatch notes) and fact-checked tweets (in days).}
	\label{fig:ccdf_social_influence} 
%    \end{minipage}
\end{figure}

\textbf{Account characteristics:} 
Fig.~\ref{fig:ccdf_social_influence} plots the CCDFs for different characteristics of Twitter accounts, namely (a) the number of followers, (b) the number of followees, and (c) the account age (in years). The plots show that tweets reported as being misleading tend to have a lower number of followers and a lower number of followees. We observe no significant differences with regard to the account age. As a further analysis, we investigated differences across verified vs. not verified accounts. Here we find that \SI{54}{\percent} of all misleading tweets were posted by verified accounts, whereas this number is \SI{61}{\percent} for not misleading tweets. %of all not misleading tweets were posted by verified accounts. %  only minor differences for misleading vs. not misleading tweets. 

\subsection{Analysis of Ratings (RQ4)}

\begin{figure}%[H]
\captionsetup{position=top}
%    \begin{minipage}[b]{0.6\textwidth}
				\centering
				% Requires \usepackage{graphicx}
				%\vspace{-.1cm}
				\subfloat[Helpfulness ratio]{\includegraphics[width=.45\linewidth]{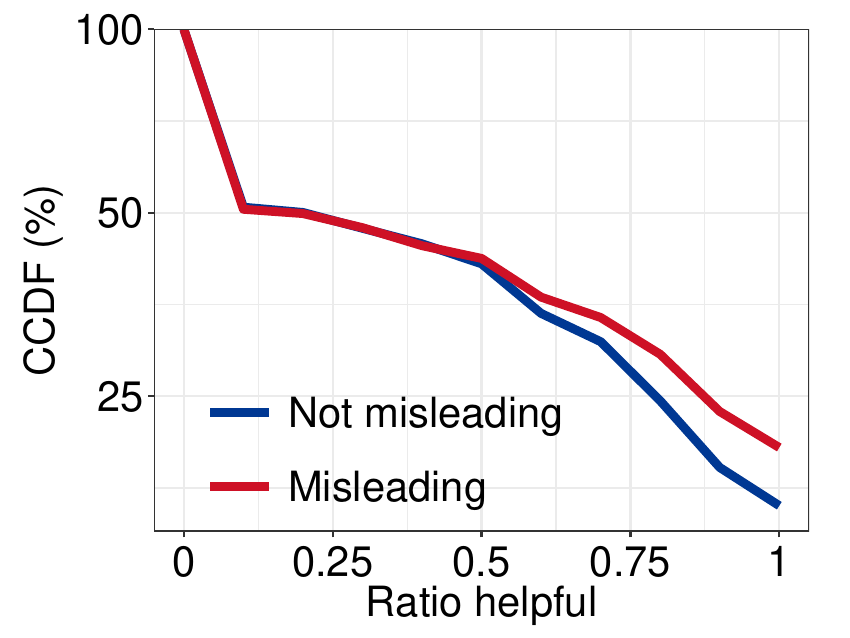}}
				\hspace{0.35cm}
				\subfloat[Total votes]{\includegraphics[width=.45\linewidth]{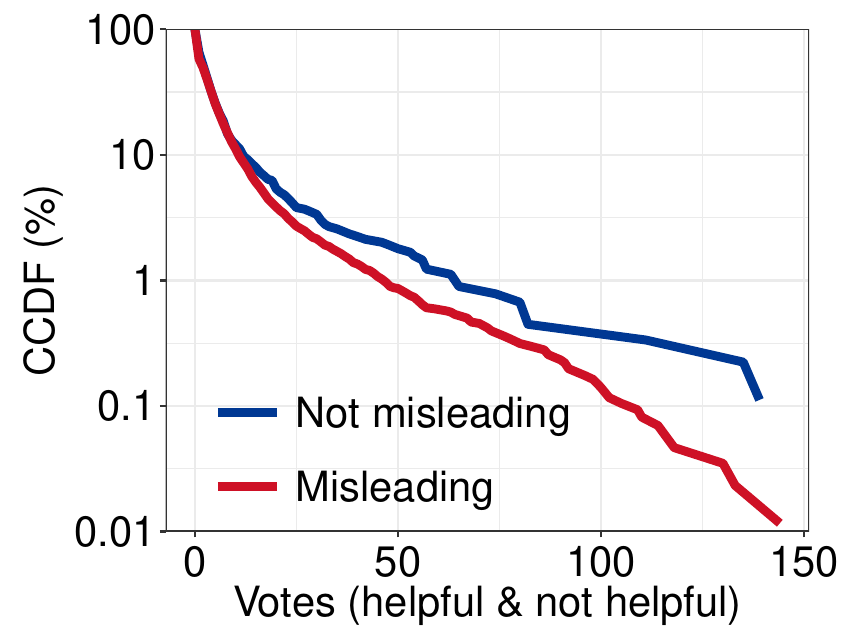}}
				%\vspace{-.1cm}
				\caption{CCDFs for (a) helpfulness ratio and (b) total votes. }
	\label{fig:ccdf_helpfulness} 
%    \end{minipage}
\end{figure}

\textbf{Helpfulness:} Fig.~\ref{fig:ccdf_helpfulness} plots the CCDFs for the helpfulness ratio ($HVotes/Votes$) and the total number of helpful and unhelpful votes ($Votes$). We find that Birdwatch notes reporting misleading tweets tend to have a higher helpfulness ratio but receive a lower number of total votes. 

\begin{figure}%[H]
	\centering
	{\includegraphics[width=\linewidth]{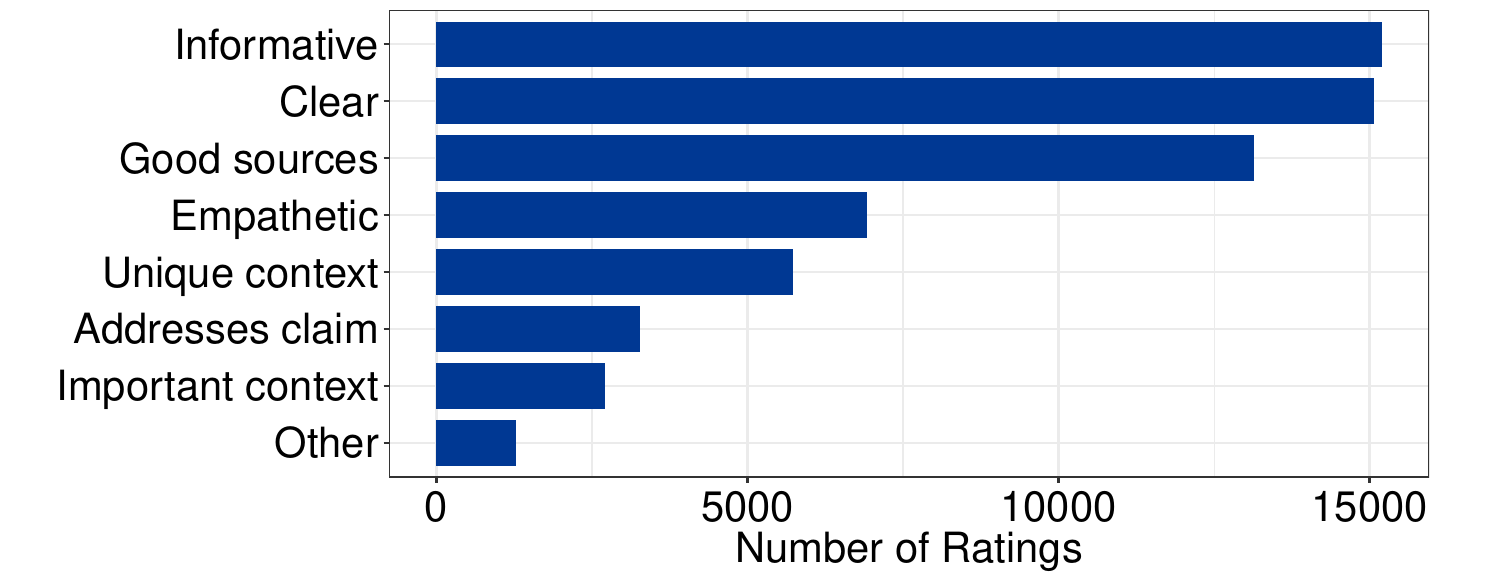}}
	\caption{Number of ratings per checkbox answer option in response to the prompt ``What about this note was helpful to you?''}
	\label{fig:reasons_helpful} 
\end{figure}

\textbf{Why do users find Birdwatch notes helpful?} Users rating a Birdwatch note helpful need to answer a checkbox question (multiple selections possible) on why they perceive it as being helpful. Fig.~\ref{fig:reasons_helpful} shows that users find that Birdwatch notes are particularly helpful if they are (i) informative (\SI{24}{\percent}), (ii) clear (\SI{24}{\percent}), and (iii) provide good sources (\SI{21}{\percent}). To a lesser extent, users also value Birdwatch notes that provide unique/informative context (\SI{14}{\percent}) and are empathetic (\SI{11}{\percent}). Only \SI{6}{\percent} of Birdwatch notes are perceived as helpful because of other reasons. 

\begin{figure}%[H]
	\centering
	{\includegraphics[width=\linewidth]{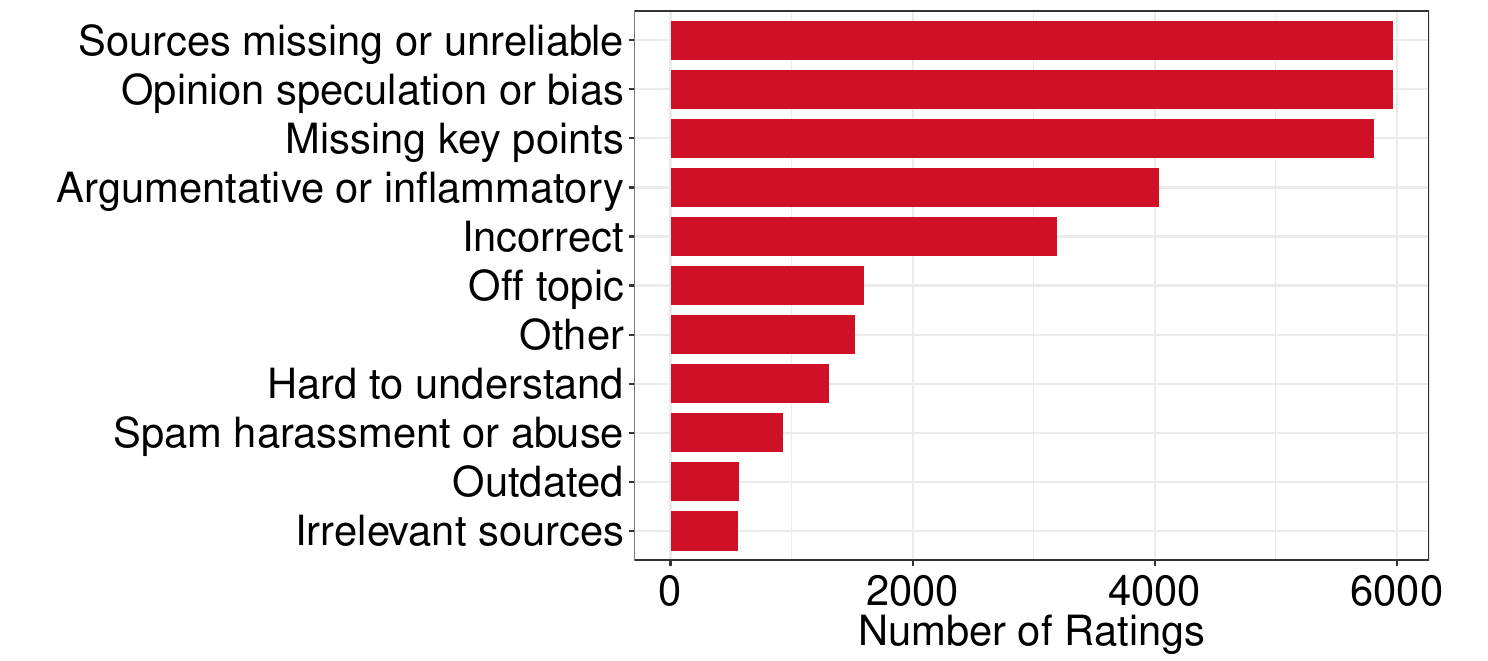}}
	\caption{Number of ratings per checkbox answer option in response to the question ``Help us understand why this note was unhelpful.''}
	\label{fig:reasons_unhelpful} 
\end{figure}

\textbf{Why do users find Birdwatch notes unhelpful?} Fig.~\ref{fig:reasons_unhelpful} shows that users find Birdwatch notes unhelpful (i) if sources are missing or unreliable (\SI{19}{\percent}), (ii) if there is opinion speculation or bias (\SI{19}{\percent}), (ii) (iii) if key points are missing (\SI{18}{\percent}), (iv) if it is argumentative or inflammatory (\SI{13}{\percent}), or (v) if it is incorrect (\SI{10}{\percent}). In some cases, users also perceive a Birdwatch note as unhelpful because it is off-topic (\SI{5}{\percent}) or hard to understand (\SI{4}{\percent}). Only few Birdwatch notes are perceived as unhelpful because of spam harassment (\SI{3}{\percent}), outdated information (\SI{2}{\percent}), irrelevant sources (\SI{2}{\percent}), or other reasons (\SI{5}{\percent}).

\subsection{Regression Model for Helpfulness (RQ4 \& RQ5)}

%\section{CCDFs: Source tweet}

We now address the question of ``what makes a Birdwatch note helpful?'' To address this question, we employ regression analyses using helpfulness as the dependent variable and (i) the user categorization, (ii) the content characteristics of the text explanation, (iii) the social influence of the source tweet as explanatory variables.

\textbf{Model specification:} Following previous research modeling helpfulness \citeeg{Lutz.2019}, we model the number of helpful votes, $HVotes$, as a binomial variable with probability parameter $\theta$ and $Votes$ trials:
%\vspace{-0.2cm}

\vspace{-.7\baselineskip}
{\scriptsize
\begin{flalign}
\logit&(\theta) =  \,\beta_0 
	+ \underbrace{\beta_1 \,\mathit{Misleading}
	+ \beta_2 \,\mathit{Trustworthy Sources}}_{\text{User categorization}} \label{eq:theta}  \\
	&+ \underbrace{\beta_3 \,\mathit{Text Complexity} 
	+ \beta_4 \,\mathit{Sentiment} 
	+ \beta_5 \,\mathit{Word Count}}_{\text{Text explanation}}  \nonumber  \\
	&+ \underbrace{\beta_6 \,\mathit{Account Age}  	 
	+ \beta_7 \,\mathit{Followers} 
	%&+ \beta_8 \,\mathit{RSentiment} 
	+ \beta_8 \,\mathit{Followees}  %\nonumber  \\
	+ \beta_{9} \,\mathit{Verified}}_{\text{Source tweet}} + \, \varepsilon ,  \nonumber  %\\ 
\end{flalign} %\vspace{-.5cm}
\begin{flalign}
HVotes \sim Binomial[Votes, \theta], & \hskip0.22\textwidth \label{eq:binom}
\end{flalign}
}
\normalsize
with intercept $\beta_0$ and error term $\varepsilon$. We estimate Eq.~\ref{eq:theta} and Eq.~\ref{eq:binom} using maximum likelihood estimation and generalized linear models. To facilitate the interpretability of our findings, we $z$-standardize all variables, so that we can compare the effects of regression coefficients on the dependent variable measured in standard deviations. 

%\textbf{Model specification for total votes: } 
We use the same set of explanatory variables as in Eq.~\ref{eq:theta} to model the total number of votes (helpful and unhelpful). The dependent variable is $Votes$, which results in the model 

\vspace{-.7\baselineskip}
{\scriptsize
\begin{flalign}
\log&({\mathup{E}(Votes \mid ^*)})  = \,\beta_0 
	+ \underbrace{\beta_1 \,\mathit{Misleading}
	+ \beta_2 \,\mathit{Trustworthy Sources}}_{\text{User categorization}}  \nonumber \\
	&+ \underbrace{\beta_3 \,\mathit{Text Complexity} 
	+ \beta_4 \,\mathit{Sentiment} 
	+ \beta_5 \,\mathit{Word Count}}_{\text{Text explanation}}  \label{eq:regression_votes}  \\
	&+ \underbrace{\beta_6 \,\mathit{Account Age}  	 
	+ \beta_7 \,\mathit{Followers} 
	%&+ \beta_8 \,\mathit{RSentiment} 
	+ \beta_8 \,\mathit{Followees}  %\nonumber  \\
	+ \beta_{9} \,\mathit{Verified}}_{\text{Source tweet}} + \, \varepsilon ,  \nonumber  %\\ 
	%+ \beta_1 \,\mathit{Misleading}
	%+ \beta_2 \,\mathit{Trustworthy Sources}  \nonumber  \\
	%&+ \beta_3 \,\mathit{Text Complexity} 
	%+ \beta_4 \,\mathit{Sentiment} 
	%+ \beta_5 \,\mathit{Word Count}  \nonumber  \\
	%&+ \beta_6 \,\mathit{Account Age}  	 
	%+ \beta_7 \,\mathit{Followers} 
	%%&+ \beta_8 \,\mathit{RSentiment} 
	%+ \beta_8 \,\mathit{Followees}  \nonumber  \\
	%& + \beta_{9} \,\mathit{Verified}
	%+ \varepsilon , \label{eq:regression_votes} 
\end{flalign}
}
\normalsize
with intercept $\beta_0$ and error term $\varepsilon$. We estimate Eq.~\ref{eq:regression_votes} via a negative binomial regression with log-transformation. The reason is that the number of votes denotes count data with overdispersion (\ie, variance larger than the mean).

\begin{figure}
    \centering
%    \begin{minipage}[b]{0.3\textwidth}
        \centering
				{\includegraphics[width=.9\linewidth]{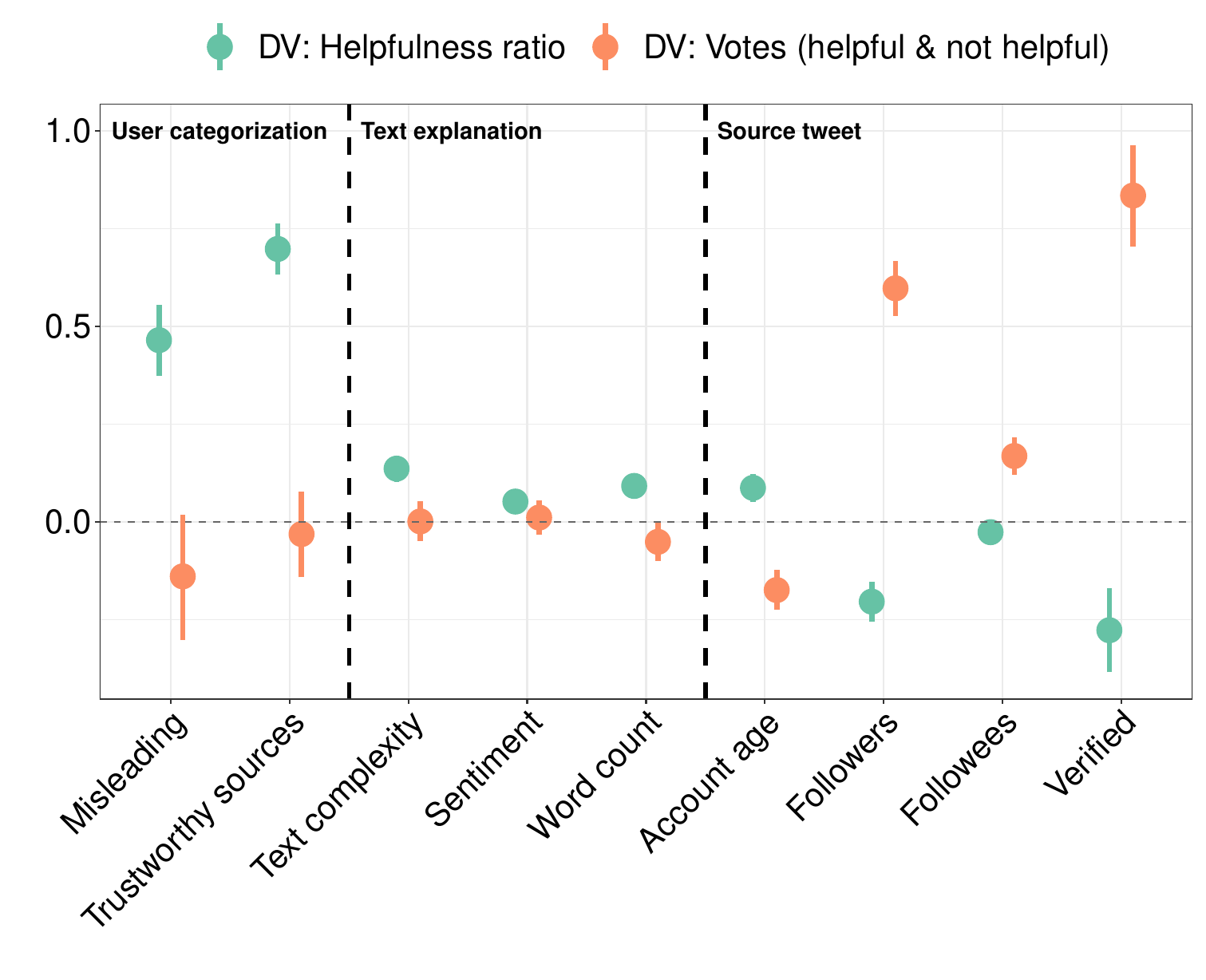}}
				%\vspace{-.1cm}
				\caption{Regression results for helpfulness ratio and total votes as dependent variables (DV). Reported are standardized parameter estimates and \SI{99}{\percent} confidence intervals.}
				\label{fig:regression_helpfulness} 
%    \end{minipage}\hspace{.05\textwidth}
\end{figure}

\textbf{Coefficient estimates:} The parameter estimates in Fig.~\ref{fig:regression_helpfulness} show that Birdwatch notes reporting misleading tweets are significantly more helpful. The coefficient for \emph{Misleading} is \num{0.465} ($p<0.001$), which implies that the odds of Birdwatch notes reporting misleading tweets are $e^{0.465} \approx 1.59$ times the odds for Birdwatch notes reporting not misleading tweets. Similarly, we find that the odds for Birdwatch notes providing trustworthy sources are 2.01 times the odds for Birdwatch notes that do not provide trustworthy sources (coef: 0.698; $p<0.001$). We also find that content characteristics of Birdwatch notes are important determinants regarding their helpfulness: the coefficients for \emph{Text complexity}, \emph{Word count}, and \emph{Sentiment} are positive and statistically significant ($p<0.001$). Hence, Birdwatch notes are estimated to be more helpful if they embed positive sentiment, and if they use more words and more complex language. For a one standard deviation increase in the explanatory variable, the estimated odds of a helpful vote increase by a factor of 1.15 for \emph{Text complexity}, 1.10 for \emph{Word count}, and 1.06 for \emph{Sentiment}. We also find that the social influence of the account that has posted the source tweet plays an important role regarding the helpfulness of Birdwatch notes. Here the largest effect sizes are estimated for \emph{Verified} and \emph{Followers} ($p<0.001$). The odds for Birdwatch notes reporting tweets from verified accounts are 0.76 times the odds for unverified accounts. A one standard deviation increase in the number of followers decreases the odds of a helpful vote by a factor of 0.81. In sum, there is a lower level of consensus for verified and high-follower accounts.  %Lastly, we also find that Birdwatch notes addressing tweets from older accounts are significantly more likely to be perceived as helpful.

We observe a different pattern for the negative binomial model with the total number of votes (helpful and unhelpful) as the dependent variable. Here we find a negative and statistically significant coefficient for \emph{Account age} ($p<0.001$) and positive and statistically significant coefficients for \emph{Verified} and \emph{Followers} ($p<0.001$). A one standard deviation increase in the \emph{Account age} is expected to decrease the total number of votes by a factor of 0.84. A one standard deviation increase in the number of followers is expected to increase the total number of votes by a factor of 1.81. Birdwatch notes reporting tweets from verified accounts are expected to receive 2.30 times more votes. %All other explanatory variables are not statistically significant at common thresholds.
Altogether, we find that the total number of votes is strongly associated with the social influence of the source tweet rather than the characteristics of the Birdwatch note itself. In contrast, the odds of a Birdwatch note to be helpful are strongly associated with both the social influence of the source tweet and the characteristics of the Birdwatch note. Birdwatch notes reported for accounts with high social influence receive more total votes but are less likely to be perceived as being helpful. 

\textbf{Analysis for subcategories of helpfulness:} We repeat our regression analysis for different subcategories of helpfulness.\footnote{For reasons of space, the coefficient estimates are relegated to the supplementary materials.} For this purpose, we estimate the effect of the explanatory variables on the frequency of the responses of Birdwatch users to the question ``What about this note was helpful/unhelpful to you?''
%The coefficient estimates %in Fig.~\ref{fig:regression_subcategories} 
%indicate that Birdwatch notes reporting misleading tweets are more informative but not necessarily more empathetic. 
We find that Birdwatch notes reporting misleading tweets are more informative but not necessarily more empathetic. 
We further observe that Birdwatch notes for misleading tweets are less likely spam harassment, opinion speculation or hard to understand. We again find that linking to trustworthy sources is crucial regarding helpfulness. Unsurprisingly, the largest effect size is observed for \emph{Trustworthy sources} on the dependent variable \emph{Good sources}. However, linking trustworthy sources also makes it more likely that a Birdwatch note is perceived to provide unique context, to be informative, to be empathetic, and to be clear. We find that Birdwatch notes that are longer, more complex, and embed a more positive sentiment are perceived as more helpful across most subcategories of helpfulness. Also, more positive sentiment is less likely to be perceived as argumentative. Our analysis further suggests reasons because of which fact checks from verified and high-follower accounts are perceived as less helpful: Birdwatch notes for these accounts are more likely to be seen as being incorrect, argumentative, opinion speculation, and spam harassment. 

\textbf{Model checks:} We conducted several checks to validate the robustness of our results: (1) We checked that variance inflation factors as an indicator of multicollinearity were below five. %This check led to the desired outcome. 
(2)~We controlled for non-linear relationships, user-specific effects, and month-level time effects. (3) We estimated separate regressions for misleading vs. not misleading tweets. In all cases, our results are robust and consistently support our findings. 

%\newpage
\section{Discussion}

%\textbf{Research Implications:} 
In this work, we provide a holistic analysis of the new Birdwatch pilot on Twitter. Our empirical findings contribute to research into misinformation and crowdsourced fact-checking. Complementary to experimental studies focusing on the question of whether crowds are \emph{able} to accurately assess social media content \citeeg{Pennycook.2019}, this work sheds light on how users \emph{interact} with community-based fact-checking systems. 

\textbf{Implications:} 
Our work has direct implications for the Birdwatch platform and future attempts to implement community-based approaches to combat misinformation on social media. We find that Twitter has developed a fact-checking platform on which  users perceive a relatively high share of community-created fact checks (Birdwatch notes) as being informative, clear and helpful. These finding are encouraging considering that earlier crowd-based fact-checking initiatives (\eg, \emph{TruthSquad}, \emph{WikiTribune}) were largely unsuccessful to produce high-quality fact-checks at scale \cite{Bhuiyan.2020}. A potential reason %for this relative success 
may be that Birdwatch differs 
%from earlier crowd-based fact-checking initiatives 
in several important ways: (i) it is directly integrated into the Twitter platform and thus has relatively low entry barriers for fact-checkers; (ii) it features a relatively sophisticated rating system to promote helpful fact-checks; (iii) it is a purely community-based approach which may reduce the problem that many users distrust professional fact-checkers \cite{Poynter.2019}; (iv) it encourages users to provide context and link to trustworthy sources, which is known to enhance credibility \cite{Zhang.2018b}.

%% Polarization 

Despite showing promising potential, our findings suggest that community-based fact-checking faces challenges with regards to social media content from influential user accounts. On Birdwatch, a large share of fact-checks is carried out for tweets from influential politicians from both sides of the political spectrum in the \US -- for which users overwhelmingly report misinformation. This indicates that users have strong interest in debunking political misinformation. We can only speculate whether tweets from these accounts are more often misleading; or rather that the fact-checking community is more likely to report posts that do not align with their political ideology. What we do find, however, is that the level of consensus between users is significantly lower if misinformation is reported for tweets from users with influential accounts. Our analysis of subcategories of helpfulness further reveals that these fact-checks are particularly likely to be perceived as being argumentative, incorrect or opinion speculation. In line with earlier research \cite{Soares.2018,Barbera.2015}, this suggests that influential user accounts (\eg, politicians) play a key role in the formation of fragmented social networks and that there is a high level of polarization for the content they produce. Polarized social media users often have vastly different interpretations of facts or even entirely different sets of acknowledged facts \cite{Otala.2021}. For crowd-based fact-checking platforms, this implies that they need to ensure independence and high diversity among fact-checkers, such that negative effects are eventually mitigated \cite{Epstein.2020}.
 
\textbf{Fact-checking guidelines:} Our findings may help to formulate guidelines for users on how to write more helpful fact checks. We find that it is crucial that users provide \emph{context} in their text explanations. Users should thus be encouraged to link trustworthy sources in their fact-checks. Furthermore, we show that content characteristics matter. Community-created fact-checks are estimated to be more helpful if they embed more positive vs. negative emotions. Users should thus avoid employing inflammatory language. We also find that community-created fact-checks that use more words and complex language are perceived as more helpful. It should thus be encouraged that users provide profound explanations and exhaust the full character limit to explain why they perceive a tweet as being misleading or not misleading.

\textbf{Limitations and directions for future research:} This works has a number of limitations, which can fuel future research as follows. First, future research should conduct analyses to better understand {which} groups of users engage in fact-checking and the accuracy of the community-created fact checks. While previous works suggest that collective wisdom can be more accurate than individual's assessments and even experts \cite{Bhuiyan.2020}, there are situations in which the crowd may perform worse. For example, it is an ongoing discussion whether crowds become more or less accurate in sequential decision making problems \cite{Frey.2020}. More research is thus necessary to better understand the conditions under which the wisdom of crowds can be unlocked for fact-checking. Second, it will be of utmost importance to understand the effect the Birdwatch feature has on the user behavior on the overall social media platform (\ie, Twitter). It has to be ensured that Birdwatch does not yield adverse side effects. Previous experience of Facebook with ``flags'' suggests that adding warning labels to some posts can make all other (unverified) posts seem more credible \cite{BBC.2017,Pennycook.2018}. Future research should seek to understand the effects of community-created fact checks on the diffusion of both the fact-checked tweet itself as well as other misleading information circulating on Twitter.

Finally, our inferences are limited to community-based fact-checking on Twitter's Birdwatch platform. Further research is thus necessary to analyze whether the observed patterns are generalizable to other crowd-based fact-checking platforms. Also, the characteristics of the restricted set of users participating in the current Birdwatch pilot phase may differ from the overall user base on social media platforms. For example, one may expect increased diversity for wider audiences, for which the wisdom of crowds literature suggests that crowds can perform better \cite{Becker.2017}. Notwithstanding these limitations, we believe that Birdwatch provides a unique opportunity to empirically analyze how users interact with a novel community-based fact-checking system, and that observing and understanding how users interact with such a system is the first step towards its rollout to wider audiences. 

\section{Conclusion}

Twitter recently launched Birdwatch, a community-driven approach to identify misinformation on social media by harnessing the wisdom of crowds. This study is the first to empirically analyze how users interact with this new feature. We find that encouraging users to provide {context} (\eg, by linking trustworthy sources) and to avoid the use of inflammatory language is crucial for users to perceive community-based fact checks as helpful. However, despite showing promising potential, our analysis also suggests that Birdwatch's community-driven approach faces challenges concerning opinion speculation and polarization among the user base -- in particular with regards to influential user accounts. Our findings are relevant both for the Birdwatch platform and future attempts to implement community-based approaches to combat misinformation on social media.
%%

%\begin{acks}
\section{Acknowledgments}
This study was supported by a grant from the German Research Foundation (DFG grant 492310022). 

%%
%% The next two lines define the bibliography style to be used, and
%% the bibliography file.
%\bibliographystyle{ACM-Reference-Format-no-doi-abbrv}
%\balance 
%\bibliography{literature}
\bibliographystyle{aaai_nodoi}
\bibliography{literature}

\begin{thebibliography}{}

\bibitem[\protect\citeauthoryear{Allcott and Gentzkow}{2017}]{Allcott.2017}
Allcott, H., and Gentzkow, M.
\newblock 2017.
\newblock Social media and fake news in the 2016 election.
\newblock {\em Journal of Economic Perspectives} 31(2):211--236.

\bibitem[\protect\citeauthoryear{Allen \bgroup et al\mbox.\egroup
  }{2020}]{Allen.2020}
Allen, J.; Howland, B.; Mobius, M.; Rothschild, D.; and Watts, D.~J.
\newblock 2020.
\newblock Evaluating the fake news problem at the scale of the information
  ecosystem.
\newblock {\em Science Advances} 6(14):eaay3539.

\bibitem[\protect\citeauthoryear{Allen \bgroup et al\mbox.\egroup
  }{2021}]{Allen.2020b}
Allen, J.; Arechar, A.~A.; Pennycook, G.; and Rand, D.~G.
\newblock 2021.
\newblock Scaling up fact-checking using the wisdom of crowds.
\newblock {\em Science Advances} 7(36).

\bibitem[\protect\citeauthoryear{Aral and Eckles}{2019}]{Aral.2019}
Aral, S., and Eckles, D.
\newblock 2019.
\newblock Protecting elections from social media manipulation.
\newblock {\em Science} 365(6456):858--861.

\bibitem[\protect\citeauthoryear{Bakabar}{2018}]{Bakabar.2018}
Bakabar, M.
\newblock 2018.
\newblock Crowdsourced factchecking.

\bibitem[\protect\citeauthoryear{Bakshy, Messing, and
  Adamic}{2015}]{Bakshy.2015}
Bakshy, E.; Messing, S.; and Adamic, L.~A.
\newblock 2015.
\newblock Exposure to ideologically diverse news and opinion on {F}acebook.
\newblock {\em Science} 348(6239):1130--1132.

\bibitem[\protect\citeauthoryear{Barber{\'a} \bgroup et al\mbox.\egroup
  }{2015}]{Barbera.2015}
Barber{\'a}, P.; Jost, J.~T.; Nagler, J.; Tucker, J.~A.; and Bonneau, R.
\newblock 2015.
\newblock Tweeting from left to right: Is online political communication more
  than an echo chamber?
\newblock {\em Psychological Science} 26(10):1531--1542.

\bibitem[\protect\citeauthoryear{BBC}{2017}]{BBC.2017}
BBC.
\newblock 2017.
\newblock Facebook ditches fake news warning flag.

\bibitem[\protect\citeauthoryear{Becker, Brackbill, and
  Centola}{2017}]{Becker.2017}
Becker, J.; Brackbill, D.; and Centola, D.
\newblock 2017.
\newblock Network dynamics of social influence in the wisdom of crowds.
\newblock {\em PNAS} 114(26):E5070--E5076.

\bibitem[\protect\citeauthoryear{Benoit \bgroup et al\mbox.\egroup
  }{2018}]{Benoit.2018}
Benoit, K.; Watanabe, K.; Wang, H.; Nulty, P.; Obeng, A.; M{\"u}ller, S.; and
  Matsuo, A.
\newblock 2018.
\newblock quanteda: An {R} package for the quantitative analysis of textual
  data.
\newblock {\em Journal of Open Source Software} 3(30):774.

\bibitem[\protect\citeauthoryear{Bhuiyan \bgroup et al\mbox.\egroup
  }{2020}]{Bhuiyan.2020}
Bhuiyan, M.~M.; Zhang, A.~X.; Sehat, C.~M.; and Mitra, T.
\newblock 2020.
\newblock Investigating differences in crowdsourced news credibility
  assessment: Raters, tasks, and expert criteria.
\newblock {\em Proceedings of the ACM on Human-Computer Interaction} 4:1--26.

\bibitem[\protect\citeauthoryear{Broniatowski \bgroup et al\mbox.\egroup
  }{2018}]{Broniatowski.2018}
Broniatowski, D.~A.; Jamison, A.~M.; Qi, S.; AlKulaib, L.; Chen, T.; Benton,
  A.; Quinn, S.~C.; and Dredze, M.
\newblock 2018.
\newblock Weaponized health communication: {T}witter bots and russian trolls
  amplify the vaccine debate.
\newblock {\em American Journal of Public Health} 108(10):1378--1384.

\bibitem[\protect\citeauthoryear{Castillo, Mendoza, and
  Poblete}{2011}]{Castillo.2011}
Castillo, C.; Mendoza, M.; and Poblete, B.
\newblock 2011.
\newblock Information credibility on {T}witter.
\newblock In {\em WWW}.

\bibitem[\protect\citeauthoryear{Conover \bgroup et al\mbox.\egroup
  }{2011}]{Conover.2011}
Conover, M.~D.; Ratkiewicz, J.; Francisco, M.; Gon{\c{c}}alves, B.; Menczer,
  F.; and Flammini, A.
\newblock 2011.
\newblock Political polarization on {T}witter.
\newblock In {\em ICWSM}.

\bibitem[\protect\citeauthoryear{{Del Vicario} \bgroup et al\mbox.\egroup
  }{2016}]{DelVicario.2016}
{Del Vicario}, M.; Bessi, A.; Zollo, F.; Petroni, F.; Scala, A.; Caldarelli,
  G.; Stanley, H.~E.; and Quattrociocchi, W.
\newblock 2016.
\newblock The spreading of misinformation online.
\newblock {\em PNAS} 113(3):554--559.

\bibitem[\protect\citeauthoryear{Ducci, Kraus, and
  Feuerriegel}{2020}]{Ducci.2020}
Ducci, F.; Kraus, M.; and Feuerriegel, S.
\newblock 2020.
\newblock Cascade-{LSTM}: A tree-structured neural classifier for detecting
  misinformation cascades.
\newblock In {\em KDD}.

\bibitem[\protect\citeauthoryear{Ecker, Lewandowsky, and
  Tang}{2010}]{Ecker.2010}
Ecker, U.~K.; Lewandowsky, S.; and Tang, D.~T.
\newblock 2010.
\newblock Explicit warnings reduce but do not eliminate the continued influence
  of misinformation.
\newblock {\em Memory \& Cognition} 38(8):1087--1100.

\bibitem[\protect\citeauthoryear{Epstein, Pennycook, and
  Rand}{2020}]{Epstein.2020}
Epstein, Z.; Pennycook, G.; and Rand, D.
\newblock 2020.
\newblock Will the crowd game the algorithm? using layperson judgments to
  combat misinformation on social media by downranking distrusted sources.
\newblock In {\em CHI}.

\bibitem[\protect\citeauthoryear{Florin}{2010}]{Florin.2010}
Florin, F.
\newblock 2010.
\newblock Crowdsourced fact-checking? what we learned from truthsquad.

\bibitem[\protect\citeauthoryear{Frey and van~de Rijt}{2020}]{Frey.2020}
Frey, V., and van~de Rijt, A.
\newblock 2020.
\newblock Social influence undermines the wisdom of the crowd in sequential
  decision making.
\newblock {\em Management Science}.

\bibitem[\protect\citeauthoryear{Friggeri \bgroup et al\mbox.\egroup
  }{2014}]{Friggeri.2014}
Friggeri, A.; Adamic, L.~A.; Eckles, D.; and Cheng, J.
\newblock 2014.
\newblock Rumor cascades.
\newblock In {\em ICWSM}.

\bibitem[\protect\citeauthoryear{Geeng, Yee, and Roesner}{2020}]{Geeng.2020}
Geeng, C.; Yee, S.; and Roesner, F.
\newblock 2020.
\newblock Fake news on {F}acebook and {T}witter: Investigating how people
  (don't) investigate.
\newblock In {\em CHI}.

\bibitem[\protect\citeauthoryear{Godel \bgroup et al\mbox.\egroup
  }{2021}]{Godel.2021}
Godel, W.; Sanderson, Z.; Aslett, K.; Nagler, J.; Bonneau, R.; Persily, N.; and
  Tucker, J.~A.
\newblock 2021.
\newblock Moderating with the mob: Evaluating the efficacy of real-time
  crowdsourced fact-checking.
\newblock {\em Journal of Online Trust and Safety} 1(1):1--36.

\bibitem[\protect\citeauthoryear{Grinberg \bgroup et al\mbox.\egroup
  }{2019}]{Grinberg.2019}
Grinberg, N.; Joseph, K.; Friedland, L.; Swire-Thompson, B.; and Lazer, D.
\newblock 2019.
\newblock Fake news on {T}witter during the 2016 {U.S.} presidential election.
\newblock {\em Science} 363(6425):374--378.

\bibitem[\protect\citeauthoryear{Gunning}{1968}]{Gunning.1968}
Gunning, R.
\newblock 1968.
\newblock {\em The Technique of Clear Writing}.
\newblock New York: McGraw-Hill.

\bibitem[\protect\citeauthoryear{Hassan \bgroup et al\mbox.\egroup
  }{2017}]{Hassan.2017}
Hassan, N.; Arslan, F.; Li, C.; and Tremayne, M.
\newblock 2017.
\newblock Toward automated fact-checking: Detecting check-worthy factual claims
  by claimbuster.
\newblock In {\em KDD}.

\bibitem[\protect\citeauthoryear{Kahan}{2017}]{Kahan.2017}
Kahan, D.~M.
\newblock 2017.
\newblock Misconceptions, misinformation, and the logic of identity-protective
  cognition.
\newblock {\em SSRN}.

\bibitem[\protect\citeauthoryear{Kim and Dennis}{2019}]{Kim.2019}
Kim, A., and Dennis, A.~R.
\newblock 2019.
\newblock Says who? {T}he effects of presentation format and source rating on
  fake news in social media.
\newblock {\em MIS Quarterly} 43(3):1025--1039.

\bibitem[\protect\citeauthoryear{Kwon and Cha}{2014}]{Kwon.2014}
Kwon, S., and Cha, M.
\newblock 2014.
\newblock Modeling bursty temporal pattern of rumors.
\newblock In {\em ICWSM}.

\bibitem[\protect\citeauthoryear{Lazer \bgroup et al\mbox.\egroup
  }{2018}]{Lazer.2018}
Lazer, D. M.~J.; Baum, M.~A.; Benkler, Y.; Berinsky, A.~J.; Greenhill, K.~M.;
  Menczer, F.; Metzger, M.~J.; Nyhan, B.; Pennycook, G.; Rothschild, D.;
  Schudson, M.; Sloman, S.~A.; Sunstein, C.~R.; Thorson, E.~A.; Watts, D.~J.;
  and Zittrain, J.~L.
\newblock 2018.
\newblock The science of fake news.
\newblock {\em Science} 359(6380):1094--1096.

\bibitem[\protect\citeauthoryear{Levy}{2021}]{Levy.2021}
Levy, R.
\newblock 2021.
\newblock Social media, news consumption, and polarization: Evidence from a
  field experiment.
\newblock {\em American Economic Review} 111(3):831--70.

\bibitem[\protect\citeauthoryear{Luca and Zervas}{2016}]{Luca.2016}
Luca, M., and Zervas, G.
\newblock 2016.
\newblock Fake it till you make it: Reputation, competition, and {Y}elp review
  fraud.
\newblock {\em Management Science} 62(12):3412--3427.

\bibitem[\protect\citeauthoryear{Lutz, Pr{\"o}llochs, and
  Neumann}{2019}]{Lutz.2019}
Lutz, B.; Pr{\"o}llochs, N.; and Neumann, D.
\newblock 2019.
\newblock The longer the better? the interplay between review length and line
  of argumentation in online consumer reviews.
\newblock In {\em International Conference on Information Systems (ICIS)}.

\bibitem[\protect\citeauthoryear{Ma \bgroup et al\mbox.\egroup
  }{2016}]{Ma.2016}
Ma, J.; Gao, W.; Mitra, P.; Kwon, S.; Jansen, B.~J.; Wong, K.-F.; and Cha, M.
\newblock 2016.
\newblock Detecting rumors from microblogs with recurrent neural networks.
\newblock In {\em ICJAI}.

\bibitem[\protect\citeauthoryear{McPherson, Smith-Lovin, and
  Cook}{2001}]{McPherson.2001}
McPherson, M.; Smith-Lovin, L.; and Cook, J.~M.
\newblock 2001.
\newblock Birds of a feather: Homophily in social networks.
\newblock {\em Annual Review of Sociology} 27:415--444.

\bibitem[\protect\citeauthoryear{Micallef \bgroup et al\mbox.\egroup
  }{2020}]{Micallef.2020}
Micallef, N.; He, B.; Kumar, S.; Ahamad, M.; and Memon, N.
\newblock 2020.
\newblock The role of the crowd in countering misinformation: A case study of
  the covid-19 infodemic.
\newblock In {\em International Conference on Big Data}.

\bibitem[\protect\citeauthoryear{Mohammad and Turney}{2013}]{Mohammad.2013}
Mohammad, S.~M., and Turney, P.~D.
\newblock 2013.
\newblock Crowdsourcing a word-emotion association lexicon.
\newblock {\em Computational Intelligence} 29(3):436--465.

\bibitem[\protect\citeauthoryear{Moravec, Minas, and
  Dennis}{2019}]{Moravec.2019}
Moravec, P.~L.; Minas, R.~K.; and Dennis, A.
\newblock 2019.
\newblock Fake news on social media: People believe what they want to believe
  when it makes no sense at all.
\newblock {\em MIS Quarterly} 43(4):1343--1360.

\bibitem[\protect\citeauthoryear{Oh, Agrawal, and Rao}{2013}]{Oh.2013}
Oh, O.; Agrawal, M.; and Rao, H.~R.
\newblock 2013.
\newblock Community intelligence and social media services: A rumor theoretic
  analysis of tweets during social crises.
\newblock {\em MIS Quarterly} 37(2):407--426.

\bibitem[\protect\citeauthoryear{Okoli \bgroup et al\mbox.\egroup
  }{2014}]{Okoli.2014}
Okoli, C.; Mehdi, M.; Mesgari, M.; Nielsen, F.~{\AA}.; and Lanam{\"a}ki, A.
\newblock 2014.
\newblock Wikipedia in the eyes of its beholders: A systematic review of
  scholarly research on {W}ikipedia readers and readership.
\newblock {\em Journal of the Association for Information Science and
  Technology} 65(12):2381--2403.

\bibitem[\protect\citeauthoryear{O'Riordan \bgroup et al\mbox.\egroup
  }{2019}]{Oriordan.2019}
O'Riordan, S.; Kiely, G.; Emerson, B.; and Feller, J.
\newblock 2019.
\newblock Do you have a source for that? understanding the challenges of
  collaborative evidence-based journalism.
\newblock In {\em OpenSym}.

\bibitem[\protect\citeauthoryear{Otala \bgroup et al\mbox.\egroup
  }{2021}]{Otala.2021}
Otala, J.; Kurtic, G.; Grasso, I.; Liu, Y.; Matthews, J.; and Madraki, G.
\newblock 2021.
\newblock Political polarization and platform migration: A study of {Parler}
  and {Twitter} usage by {United States of America Congress} members.
\newblock In {\em WWW Companion}.

\bibitem[\protect\citeauthoryear{Pennycook and Rand}{2019a}]{Pennycook.2019}
Pennycook, G., and Rand, D.~G.
\newblock 2019a.
\newblock Fighting misinformation on social media using crowdsourced judgments
  of news source quality.
\newblock {\em PNAS} 116(7):2521--2526.

\bibitem[\protect\citeauthoryear{Pennycook and Rand}{2019b}]{Pennycook.2019b}
Pennycook, G., and Rand, D.~G.
\newblock 2019b.
\newblock Lazy, not biased: Susceptibility to partisan fake news is better
  explained by lack of reasoning than by motivated reasoning.
\newblock {\em Cognition} 188:39--50.

\bibitem[\protect\citeauthoryear{Pennycook \bgroup et al\mbox.\egroup
  }{2021}]{Pennycook.2021}
Pennycook, G.; Epstein, Z.; Mosleh, M.; Arechar, A.~A.; Eckles, D.; and Rand,
  D.~G.
\newblock 2021.
\newblock Shifting attention to accuracy can reduce misinformation online.
\newblock {\em Nature}  1--6.

\bibitem[\protect\citeauthoryear{Pennycook, Cannon, and
  Rand}{2018}]{Pennycook.2018}
Pennycook, G.; Cannon, T.~D.; and Rand, D.~G.
\newblock 2018.
\newblock Prior exposure increases perceived accuracy of fake news.
\newblock {\em Journal of Experimental Psychology: General} 147(12):1865--1880.

\bibitem[\protect\citeauthoryear{{Pew Research Center}}{2016}]{Pew.2016}
{Pew Research Center}.
\newblock 2016.
\newblock News use across social media platforms 2016.

\bibitem[\protect\citeauthoryear{Poynter}{2019}]{Poynter.2019}
Poynter.
\newblock 2019.
\newblock Most {Republicans} don't trust fact-checkers, and most {Americans}
  don’t trust the media.

\bibitem[\protect\citeauthoryear{Pr{\"o}llochs, B{\"a}r, and
  Feuerriegel}{2021a}]{Prollochs.2021b}
Pr{\"o}llochs, N.; B{\"a}r, D.; and Feuerriegel, S.
\newblock 2021a.
\newblock Emotions explain differences in the diffusion of true vs. false
  social media rumors.
\newblock {\em Scientific Reports} 11(22721).

\bibitem[\protect\citeauthoryear{Pr{\"o}llochs, B{\"a}r, and
  Feuerriegel}{2021b}]{Prollochs.2021a}
Pr{\"o}llochs, N.; B{\"a}r, D.; and Feuerriegel, S.
\newblock 2021b.
\newblock Emotions in online rumor diffusion.
\newblock {\em EPJ Data Science} 10(1):51.

\bibitem[\protect\citeauthoryear{Qazvinian \bgroup et al\mbox.\egroup
  }{2011}]{Qazvinian.2011}
Qazvinian, V.; Rosengren, E.; Radev, D.~R.; and Mei, Q.
\newblock 2011.
\newblock Rumor has it: Identifying misinformation in microblogs.
\newblock In {\em EMNLP}.

\bibitem[\protect\citeauthoryear{Rinker}{2019}]{Rinker.2019}
Rinker, T.~W.
\newblock 2019.
\newblock {\em {sentimentr}: Calculate Text Polarity Sentiment}.
\newblock Buffalo, New York.
\newblock version 2.7.1.

\bibitem[\protect\citeauthoryear{Shao \bgroup et al\mbox.\egroup
  }{2016}]{Shao.2016}
Shao, C.; Ciampaglia, G.~L.; Flammini, A.; and Menczer, F.
\newblock 2016.
\newblock Hoaxy: A platform for tracking online misinformation.
\newblock In {\em WWW Companion}.

\bibitem[\protect\citeauthoryear{Soares, Recuero, and Zago}{2018}]{Soares.2018}
Soares, F.~B.; Recuero, R.; and Zago, G.
\newblock 2018.
\newblock Influencers in polarized political networks on {T}witter.
\newblock In {\em International Conference on Social Media and Society}.

\bibitem[\protect\citeauthoryear{Starbird}{2017}]{Starbird.2017}
Starbird, K.
\newblock 2017.
\newblock Examining the alternative media ecosystem through the production of
  alternative narratives of mass shooting events on {T}witter.
\newblock In {\em ICWSM}.

\bibitem[\protect\citeauthoryear{Vo and Lee}{2018}]{Vo.2018}
Vo, N., and Lee, K.
\newblock 2018.
\newblock The rise of guardians: Fact-checking url recommendation to combat
  fake news.
\newblock In {\em SIGIR}.

\bibitem[\protect\citeauthoryear{Vosoughi, Mohsenvand, and
  Roy}{2017}]{Vosoughi.2017}
Vosoughi, S.; Mohsenvand, M.~N.; and Roy, D.
\newblock 2017.
\newblock Rumor gauge: Predicting the veracity of rumors on {T}witter.
\newblock {\em ACM Transactions on Knowledge Discovery from Data} 11(4):1--36.

\bibitem[\protect\citeauthoryear{Vosoughi, Roy, and Aral}{2018}]{Vosoughi.2018}
Vosoughi, S.; Roy, D.; and Aral, S.
\newblock 2018.
\newblock The spread of true and false news online.
\newblock {\em Science} 359(6380):1146--1151.

\bibitem[\protect\citeauthoryear{Woolley \bgroup et al\mbox.\egroup
  }{2010}]{Woolley.2010}
Woolley, A.~W.; Chabris, C.~F.; Pentland, A.; Hashmi, N.; and Malone, T.~W.
\newblock 2010.
\newblock Evidence for a collective intelligence factor in the performance of
  human groups.
\newblock {\em Science} 330(6004):686--688.

\bibitem[\protect\citeauthoryear{Wu \bgroup et al\mbox.\egroup
  }{2019}]{Wu.2019}
Wu, L.; Morstatter, F.; Carley, K.~M.; and Liu, H.
\newblock 2019.
\newblock Misinformation in social media: definition, manipulation, and
  detection.
\newblock {\em SIGKDD Explorations Newsletter} 21(2):80--90.

\bibitem[\protect\citeauthoryear{Zhang \bgroup et al\mbox.\egroup
  }{2018}]{Zhang.2018b}
Zhang, A.~X.; Ranganathan, A.; Metz, S.~E.; Appling, S.; Sehat, C.~M.; Gilmore,
  N.; Adams, N.~B.; Vincent, E.; Lee, J.; Robbins, M.; et~al.
\newblock 2018.
\newblock A structured response to misinformation: Defining and annotating
  credibility indicators in news articles.
\newblock In {\em WWW Companion}.

\end{thebibliography}

%%
%% If your work has an appendix, this is the place to put it.
%\appendix

%\clearpage
%\appendix
%
%\onecolumn
%
%
%\begin{center}
%\Huge
%\textbf{Supplementary Materials}
%\end{center}
%\normalsize
%
%\vspace{1cm}
%
%\section{Regression Results for Subcategories of Helpfulness}
%\renewcommand\thefigure{A.\arabic{figure}}   
%\setcounter{figure}{0}
%
%\begin{figure*}%[H]
%\captionsetup{position=top}
%%    \begin{minipage}[b]{0.6\textwidth}
				%\centering
				%% Requires \usepackage{graphicx}
				%%\vspace{-.1cm}
				%\subfloat[Subcategories: Helpful]{\includegraphics[width=.95\linewidth]{figures/regression_coef_helpful_subcategories}\label{fig:regression_subcategories_helpful}}
				%%\hspace{0.35cm}
				%\\
				%\subfloat[Subcategories: Unhelpful]{\includegraphics[width=.95\linewidth]{figures/regression_coef_notHelpful_subcategories}\label{fig:regression_subcategories_NotHelpful}}
				%%\vspace{-.1cm}
				%\caption{Regression results for subcategories of (a) helpfulness and (b) unhelpfulness as dependent variables (DV). Reported are standardized parameter estimates and \SI{95}{\percent} confidence intervals.}
	%\label{fig:regression_subcategories} 
%%    \end{minipage}
%\end{figure*}
%\clearpage

\end{document}